\documentclass[useAMS,usenatbib]{mn2e}

\usepackage{graphics,graphicx}
\usepackage{paralist}
\usepackage{amsmath}
\usepackage{amssymb}
\usepackage{lscape}
\usepackage{rotating}
\usepackage{afterpage}

\title[HMXBs in star-forming galaxies]{X-ray emission from star-forming galaxies - I. High-mass X-ray binaries}

\author[S. Mineo, M. Gilfanov and R. Sunyaev]{S. Mineo$^{1}$\thanks{E-mail:
mineo@mpa-garching.mpg.de}, M. Gilfanov $^{1, 2}$ and R. Sunyaev $^{1, 2}$\\
$^{1}$Max Planck Institut f\"ur Astrophysik, Karl-Schwarzschild-Str. 1 85741 Garching, Germany\\
$^{2}$Space Research Institute of Russian Academy of Sciences, Profsoyuznaya 84/32, 117997 Moscow, Russia}
\begin{document}
\sloppypar

\date{Accepted. Received; in original form}

\pagerange{\pageref{firstpage}--\pageref{lastpage}} \pubyear{2011}

\maketitle

\label{firstpage}

\begin{abstract}
Based on a homogeneous set of X-ray, infrared and ultraviolet observations from \textit{Chandra}, Spitzer,  GALEX and 2MASS archives, we study  populations of high-mass X-ray binaries  (HMXBs) in a sample of 29 nearby star-forming galaxies and their relation with the star formation rate (SFR). In agreement with previous results, we find that HMXBs are a good tracer of the recent star formation activity in the host galaxy and their collective luminosity and number scale with  the SFR, in particular,  $L_\rmn{X}\approx 2.6\cdot 10^{39} \times \rmn{SFR}$.  However, the scaling relations still bear a rather large dispersion of $rms \sim 0.4$ dex, which we believe is of a physical origin.  

We present the catalog of 1057  X-ray sources detected within the $D25$ ellipse for galaxies of our sample and construct the average X-ray luminosity function (XLF) of HMXBs with substantially improved statistical accuracy and better control of systematic effects  than achieved in previous studies. The XLF follows a power law with slope of 1.6 in the $\log(L_\rmn{X})\sim 35-40$ luminosity range with a moderately  significant evidence for a break or cut-off at $L_{\rmn{X}} \sim 10^{40}\, \rmn{erg}/\rmn{s}$. As before, we did not find any features at the Eddington limit for a neutron star or a stellar mass black hole.

We discuss implications of our results for the theory of binary evolution. In particular we estimate  the fraction of compact objects that once upon their lifetime experienced an X-ray active phase powered by accretion from a high mass companion and obtain a rather large number, $f_{\rmn{X}}\sim 0.2\times \left(0.1~{\rm Myr}/\tau_{\rmn{X}}  \right)$ ($\tau_{\rmn{X}}$ is the life time of the X-ray active phase). This is $\sim 4$ orders of magnitude more frequent than in LMXBs. We also derive constrains on the mass distribution of the secondary star in HMXBs.  
\end{abstract}

\begin{keywords}
X-rays: binaries -- X-rays: galaxies -- galaxies: starburst -- galaxies: spirals -- galaxies: irregulars -- stars: formation.
\end{keywords}

\section{Introduction}

The recent star formation rate of a galaxy is primarily measured from the light emitted by massive stars: given their short lifetimes, their collective luminosity is, to the first approximation, proportional to the rate at which they are currently being formed.
The scale factors  relating luminosities at different wavelengths and SFR are usually derived from stellar population synthesis models \citep[see review by][]{1998ARA&A..36..189K}. Most of the conventional SFR indicators are known to be subject to various uncertainties, one of which is caused by the interstellar extinction. 

An alternative  method of SFR determination is based on the X-ray emission of a galaxy \citep[e.g.][]{1978ComAp...7..183S, 2002AJ....124.2351B, 2003MNRAS.339..793G, 2003A&A...399...39R}. Indeed,  X-ray output of a normal gas-poor galaxy is  dominated by the collective emission of its X-ray binary (XRB) population \citep[see review by][]{2006ARA&A..44..323F}. As the evolutionary time scale of high-mass X-ray binaries does not exceed $\sim{\rm few}\times 10^7$ yrs, their X-ray emission is nearly prompt \citep{2007AstL...33..437S}. That makes them a potentially good tracer of the recent star formation activity in the host galaxy. However, in distant galaxies, emission from HMXBs can not be separated from that  of low-mass X-ray binaries (LMXBs) having much longer evolutionary time scales and therefore unrelated to recent star-formation or from the contribution of the supermassive black  hole. For this reason the X-ray based method is suitable for young normal galaxies with sufficiently large value of the specific, per unit stellar mass, SFR, the X-ray output of which is dominated by HMXBs. Due to large penetrating power of X-rays, the X-ray based method is much less affected by absorption than conventional indicators. This makes it an important tool for cross-calibration of different SFR proxies and for star-formation diagnostics in galaxies, especially in the distant ones.  

Although the theoretical idea was formulated in the 70-ies \citep{1978ComAp...7..183S} and the first supporting evidence was obtained in 90-ies \citep{1992ApJ...388...82D}, the full observational justification and calibration of the method required outstanding angular resolution and could be achieved only with advent of Chandra. 
During the decade following its launch several authors have discussed the X-ray emission of compact sources and its relation to the SFR of the host galaxy \citep[][and several others]{2003MNRAS.339..793G, 2003A&A...399...39R, 2007A&A...463..481P, 2010ApJ...724..559L}.
All these works have established a general agreement regarding the correlation of the X-ray luminosity with SFR, however there was some discrepancy between the scale factors derived by different authors. In addition, a notable dispersion in the relation was observed.
One of the limiting factors of previous studies was the heterogeneous nature of the X-ray and SFR data. With the large numbers of galaxies now available in \textit{Chandra}, Spitzer and GALEX public data archives it has become possible to construct large samples of galaxies for which homogeneous sets of multi-wavelength measurements are available. This motivated us to revisit the problem of the $L_{\rmn{X}}-\rmn{SFR}$ relation in order to obtain its more accurate calibration and understand  the origin of the dispersion seen in previous studies. 

Study of the populations of HMXBs and its dependence on the  metallicity and other properties  of the host galaxy is of considerable interest on its own. The specific luminosity and numbers of sources as well as their luminosity distribution can be used for verification and calibration of the binary evolution and population synthesis models. \citet{2003MNRAS.339..793G} showed that in a wide range of SFR, the luminosity distribution of HMXBs in a galaxy can be approximately described by a universal luminosity function, which is a power law with slope of $\sim 1.6$, whose normalization is proportional to the SFR.  They searched for features in the averaged XLF of nearby galaxies, corresponding to the Eddington luminosities of neutron stars and black holes, but did not find any.  Among other reasons, of a more physical nature, they also considered
the possibility of various systematic effects (e.g. distance uncertainties) which could smear and dilute the expected features.  A larger sample of galaxies with well controlled systematic uncertainties may allow a substantial progress in this direction.

The present paper is the first of a series of publications investigating  X-ray emission from star-forming galaxies. It will  focus on 
the study of the population of HMXBs and its relation to the star formation rate of the host galaxy.
The structure of the paper is as follows. In Section 2 we present the selection criteria and the resulting sample of galaxies. The X-ray data analysis methods are described In Section 3. In Section 4 we study the spatial distribution of compact sources in our sample galaxies in order to minimize the contamination by LMXBs, cosmic X-ray background (CXB) and active galactic nuclei (AGN). The analysis of far-infrared, near-infrared and ultraviolet data is unfold in Section 5. In  Section 6 we present the star formation rate estimators that we use in the present work. 
In Section 7 we study luminosity distributions of HMXBs in individual galaxies and their average XLF.
In Section 8 we investigate the behavior of the collective luminosity of HMXBs and its dependence on the star-formation rate.
In Section 9 and 10  we discuss astrophysical implications of the obtained results and various aspects of their application.
In Section 11 we summarize our findings.

\section{The sample}
\label{sec:sample}

\subsection{Primary sample of resolved galaxies}

\afterpage{
\clearpage
\begin{landscape}
\begin{table}
\begin{minipage}{230mm}
\centering
\caption{Primary sample: galaxies resolved by \textit{Chandra}.}
\label{table:sample1}
\begin{tabular}{@{}l c l c c c c c c c c c c c c@{}}
\hline
Galaxy &  $D$ & Ref.\footnote{References for distances: (1) \citet{2003AJ....126.1607S}, (2) \citet{1988Sci...242..310T}, (3) \citet{2008A&A...486..151G}, (4) \citet{2006A&A...452..141G}, (5) \citet{2003Ap.....46..144M}, (6) \citet{1997ApJS..109..333W}, (7) \citet{2006ApJS..165..108S}, (8) \citet{1999ApJ...526..599S}, (9) \citet{2005JKAS...38....7I}, (10) \citet{2004AJ....127..660S}, (11) \citet{2002AJ....124..811D}, (12) \citet{2005AJ....129.1331S}, (13) \citet{2006MNRAS.372.1735T}, (14) \citet{2004ApJ...608...42S}, (15) \citet{2000A&AS..142..425D}, (16) \citet{2009AJ....138..323T}, (17) \citet{2004A&A...426..787W}.} & Hubble type & $i$\footnote{Inclination angles taken from \citet{1988Sci...242..310T}.} & $R_{\rmn{X}}$\footnote{Radius of the elliptical galactic region that was used for the  analysis (see Sect.\ref{sec:prof_cxb}).} & SFR\footnote{Star formation rate from Spitzer and GALEX data (see Sect. \ref{sec:sfr}).} & $M_{\star}$\footnote{Stellar mass from 2MASS data (see Sect. \ref{sec:multiwav}).}  & $\rmn{SFR}/M_{\star}$ &  $\log L_{\rmn{lim}}$\footnote{Sensitivity limit in decimal logarithmic units, defined as the luminosity level corresponding to a value of the incompleteness function $K(L)=0.6$.} & $N_{\rmn{XRB}}$\footnote{Number of XRBs with luminosity above the sensitivity limit, detected within the galactic region where the HMXBs dominate the compact X-ray source population (Sect. \ref{sec:spatial}).} & $\log L_{\rmn{XRB}}$\footnote{Base-10 logarithm of the total X-ray luminosity of galaxies in the $0.5-8\,\rmn{keV}$ band, estimated by using eq. (\ref{eq:lx}).} & $N_{\rmn{CXB}}$\footnote{Predicted number CXB sources within the selected region (see Sect. \ref{sec:spatial}), with luminosity above the sensitivity limit.}& $\log L_{\rmn{CXB}}$\footnote{Predicted contribution of CXB sources to the total X-ray luminosity of galaxies.} \\
       &  (Mpc)    &    &     &  (deg) & (arcmin) & ($M_{\odot}$ yr$^{-1}$) & ($10^{10}\,M_{\odot}$) &  ($10^{-10}$ yr$^{-1}$) &  $(\rmn{erg}\,\rmn{s}^{-1})$ & & $(\rmn{erg}\,\rmn{s}^{-1})$ & & $ (\rmn{erg}\,\rmn{s}^{-1})$\\
\hline
\hline
NGC~0278       & $11.8^{+2.4}_{-2.0}$ & (2) & SAB(rs)b     & 0 & 1.0 & 4.1		  & $0.7$ & 5.7 & 37.22 & 10 & 39.34 & 1.6  & 38.38\\
NGC~0520       & $27.8^{+5.6}_{-4.7}$  & (2) & Pec	      & 69 & 1.8 & 11.6		  & 4.7	 	       & 2.4  & 38.18 & 11 & 40.40 & 2.3  & 39.39\\	
NGC~1313       & $4.1\pm0.2$ & (3) & SB(s)d	      & 38 & 2.0 & 0.44		  & 0.1	 	       & 4.2  & 36.82 & 8  & 39.69 & 2.4  & 37.92\\	
NGC~1427A      & $15.9\pm1.6$ & (4) & IB(s)m       & 55 & 1.2 & 0.21		  & $7.3\times10^{-3}$ & 28.7 & 37.72 & 6  & 39.72 & 1.0  & 38.59\\	
NGC~1569       & $1.9\pm 0.2$ & (5) & IB 	      & 64 & 1.0 & $7.8\times10^{-2}$  & $2.8\times10^{-2}$ & 2.8  & 35.71 & 10 & 37.93 & 0.78 & 36.52\\	
NGC~1672       & 16.8 & (1) & SB(s)b	      & 37 & 2.5 & 12.0		  & 4.6		       & 2.6  & 37.74 & 25 & 40.38 & 7.1  & 39.45\\ 	
NGC~2139       & $26.7^{+5.8}_{-4.8}$ & (6) & SAB(rs)cd    & 53 & 1.4 & 3.8		  & 0.91	       & 4.2  & 38.14 & 15 & 40.47 & 1.8  & 39.26\\	
NGC~2403       & $3.1\pm 0.2$ & (7) & SAB(s)cd     & 62 & 4.5 & 0.52		  & 0.27 	       & 1.9  & 36.16 & 42 & 39.37 & 17.7 & 38.31\\	
NGC~3034 (M82) & $3.9\pm 0.3$ & (8) & I0 	      & 66 & 5.0 & 10.5		  & 2.4		       & 4.3  & 36.86 & 54 & 39.86 & 6.5  & 38.37\\	
NGC~3079       & 18.2 & (1) & SB(s)c	      & 88 & 1.9 & 6.0		  & 4.0		       & 1.5  & 37.98 & 14 & 39.79 & 0.63 & 38.57\\	
NGC~3310       & 19.8 & (1) & SAB(r)bc pec & 25 & 1.4 & 7.1		  & 0.98	       & 7.2  & 37.84 & 23 & 40.78 & 2.1  & 39.03\\	
NGC~3556       & $10.7^{+1.9}_{-1.6}$ & (16) & SB(s)cd      & 81 & 3.7 & 	3.1	  & 	1.7	       & 1.8  & 37.27 & 30 & 39.79 & 5.0 & 38.86\\	
NGC~3631       & $24.3\pm 1.6$ & (9) & SA(s)c	      & 36 & 2.4 & 4.6		  & 2.6		       & 1.8  & 37.92 & 29 & 40.60 & 8.6  & 39.77\\	
NGC~4038/39    & $13.8\pm 1.7$  & (10) & - & - & 2.3\footnote{for The Antennae we have analyzed a region that includes the two galaxies in the merging system (NGC~4038 and NGC~4039). It is centered at $\alpha_{\rmn{J2000}} = 180.472$ deg and $\delta_{\rmn{J2000}}= -18.881$ deg and has axis ratio=1.4 and PA=78 deg.} & 5.4		  & 3.1		       & 1.7  & 36.92 & 83 & 40.23 & 12.4 & 39.15\\	
NGC~4194       & $39.1^{+7.9}_{-6.6}$ & (2) & IBm pec      & 49 & 0.9 & 16.8		  & 2.1		       & 8.0  & 38.64 & 4  & 40.45 & 0.5  & 39.12\\	
NGC~4214       & $2.5\pm 0.3$ & (11) & IAB(s)m      & 37 & 2.5 & 0.17		  & $3.3\times10^{-2}$ & 5.2  & 36.19 & 14 & 38.33 & 5.1  & 37.71\\	
NGC~4490       & $7.8^{+1.6}_{-1.3}$& (2) & SB(s)d pec   & 65 & 3.4 & 1.8		  & 0.39	       & 4.7  & 37.10 & 32 & 40.05 & 2.8  & 38.17\\	
NGC~4625       & $8.2^{+1.7}_{-1.4}$ & (2) & SAB(rs)m pec & 23 & 0.6 & 0.09		  & $7.0\times10^{-2}$ & 1.3  & 37.01 & 4  & 37.98 & 0.57 & 37.67\\	
NGC~4631       & $7.6\pm 0.1$ & (12) & SB(s)d	      & 85 & 5.8 & 4.0		  & 1.4		       & 2.9  & 37.09 & 23 & 39.68 & 7.0  & 38.78\\	
NGC~5194 (M51A)& $7.6\pm 1.0 $& (13) & SA(s)bc pec  & 64 & 5.8 & 3.7		  & 2.8		       & 1.3  & 37.05 & 69 & 39.93 & 26.4 & 39.33\\	
NGC~5253       & $4.1\pm 0.5$ & (7) & Im pec	      & 77 & 2.3 & 0.38		  & $5.9\times10^{-2}$ & 6.5  & 36.07 & 17 & 38.52 & 5.0  & 37.82\\	
NGC~5457 (M101)& $6.7\pm 0.3$ & (14) & SAB(rs)cd    & 0 & 4.0 & 1.5		  & 0.99	       & 1.5  & 36.36 & 96 & 39.47 & 38.2 & 39.05\\	
NGC~5474       & 6.8 & (15) & SA(s)cd pec  & 37 & 1.6 & 0.18		  & $9.1\times10^{-2}$ & 2.0  & 37.04 & 10 & 38.90 & 2.5  & 38.27\\	
NGC~5775       & $26.7^{+11.9}_{-8.2}$ & (2) & Sb(f)	      & 84 & 2.6 & 5.3		  & 6.3		       & 0.85 & 38.01 & 25 & 40.53 & 2.5  & 39.33\\	
NGC~7090       & 7.6 & (1) & SBc	      & 90 & 3.1 & 0.29		  & 0.22	       & 1.3  & 37.23 & 11 & 39.10 & 1.5  & 38.19\\	
NGC~7541       & $34.9^{+6.6}_{-5.7}$ & (6) & SB(rs)bc pec & 75 & 1.4 & 14.7		  & 4.9		       & 3.0  & 38.41 & 13 & 40.25 & 0.81 & 39.16\\	
NGC~7793       & $4.0^{+0.7}_{-0.6}$ & (16) & SA(s)d	      & 50 & 2.0 & 0.29		  & 0.15	       & 2.0  & 36.55 & 9  & 38.23 & 3.2  & 37.88\\	
UGC 05720      & $24.9^{+11.1}_{-7.7}$ & (2) & Im pec       & 28 & 0.6 & 1.8		  & 0.37	       & 4.9  & 38.36 & 4  & 39.68 & 0.18 & 38.37\\	
CARTWHEEL      & 122.8 & (17) & S pec (Ring) & - & 0.7 & 17.6                & 4.2	       & 4.2  & 39.20 & 11 & 41.23 & 0.68 & 40.01\\	
\hline
\end{tabular}
\end{minipage}
\end{table}
\end{landscape}
}

We started to build the sample of star-forming galaxies from the entire "Normal 
Galaxies" section of the \textit{Chandra} Data Archive. It includes over 300 galaxies of all
morphological types. We based our selection on the following four criteria.

\begin{inparaenum}[\itshape i\upshape)]
\item \textit{Hubble type:} we selected only late-type galaxies, i.e. those having numerical index of stage along the Hubble sequence $T>0$ \citep{1976srcb.book.....D}. According to this definition, $S0$ galaxies were excluded since the origin of their far-infrared and ultraviolet emission is not 
associated with recent star formation. This first cut reduced the sample to  172 spiral and irregular galaxies.

\item \textit{Specific SFR:} as  star-forming galaxies contain also LMXBs, unrelated to the recent star-formation activity, the contribution of the latter will contaminate the $L_{\rmn{X}}-$SFR relation. Although separation of the high- and low- mass XRBs in external galaxies is virtually impossible, the LMXBs contribution can be controlled in a statistical way. To this end we use the fact that their population is proportional  to the stellar mass $M_{\star}$ of the host galaxy \citep{2004MNRAS.349..146G}, whereas HMXBs scale with SFR \citep{2003MNRAS.339..793G}, therefore the specific SFR ($\rmn{SFR}/M_{\star}$) can serve as a measure of their relative contributions to the X-ray luminosity of the galaxy. We estimated the stellar mass of the galaxies as it is described in
\citet{2004MNRAS.349..146G}, using the $K_{\rmn{S}}$-band (2.16$\,\umu$m) magnitudes from
The Two Micron All Sky Survey (2MASS). For the purpose of the sample selection we used IRAS data to estimated SFRs.   Based on the fluxes at 60$\,\umu$m and 100$\,\umu$m taken 
from IRAS catalogues, we first computed the $42.5-122.5\,\umu$m flux (FIR) following 
\citet{1985ApJ...298L...7H}. Adopting a standard correction factor of 1.7 we converted the latter 
into a $8-1000\,\umu$m flux and computed SFR  using the calibration of 
\citet{2003ApJ...586..794B} (eq. 4). Taking into account the shapes of luminosity distributions for high- and low-mass X-ray binaries and typical point sources detection sensitivities of Chandra data, we estimated that a threshold of $\mathrm{SFR}/M_{\star} >1 \times 10^{-10}$ yr$^{-1}$ would be a reasonable compromise, sufficient for our purpose. After this selection the number of galaxies reduced to 88.

\item \textit{Exposure time:}  we used only those observations in which the galaxy is observed sufficiently close to the aim-point, with sufficiently long exposure-time ($t_{\rmn{exp}} \geq 15$ ks). 
In case multiple observations were available in the \textit{Chandra} archive for a given galaxy, we combined them in order to improve the sensitivity limit, $L_{\rmn{lim}}$. After applying this selection criterion, 36 galaxies  are left in our sample.

\item \textit{Distance:} for the galaxies passed through the above criteria,  a number of redshift independent distances, determined with different methods, were collected from NED\footnote{NASA/IPAC Extragalactic Database}. 
Following the results of \citet{1992PASP..104..599J}, we chose the most accurate among them.  For our primary sample we chose a distance threshold of 40 Mpc, in order to spatially 
resolve the X-ray compact sources with \textit{Chandra}. This also allows us to identify and separate the central AGN from the rest of detected point sources. After this last selection, 28 galaxies are left in our sample.

\item \textit{Sensitivity} The galaxy list produced in the  result of this selection had  sensitivities in the range  of $\log(L_{\rmn{X}})\sim 35.7-38.6$ with the median value of $\log(L_{\rmn{X}})\sim 37.1$.
 As we perform an accurate incompleteness correction, no further filtering based on the point source detection sensitivity was performed. 

\end{inparaenum}

For the purpose of binary population studies, this rigorously defined sample was complemented by two more galaxies. In order to improve statistics of the high luminosity end of the XLF we added the Cartwheel galaxy.
It has the largest SFR value among our resolved galaxies and although is more distant (D$\sim$ 123 Mpc), the brightest sources, with $\log(L_{\rmn{X}})\ga 39$, do not suffer from significant sources confusion in Chandra data. 
To expand the low luminosity end we used the ASCA data of SMC observations \citep{2003PASJ...55..161Y}. The SMC data were used only for the XLF construction.

The so defined primary sample is presented in Table \ref{table:sample1}. It includes 29 galaxies, with SFR values ranging from $\sim 0.1$ to $\sim 20$ $M_{\odot}$ yr$^{-1}$.

\subsection{High-SFR sample}
In order to explore the $L_{\rmn{X}}-\rmn{SFR}$ relation in the   high SFR regime, we relaxed the distance constraint and constructed a secondary sample from the galaxies that passed  the first and second criteria,  and are characterized by high star-formation rates, $SFR\ga 15~M_\odot$/yr.  In total,  we found 8 galaxies with SFR ranging from  $15$ to $290$ $M_{\odot}$ yr$^{-1}$ (all, naturally, located outside $D>40$ Mpc). 

These galaxies were treated as spatially unresolved in our analysis (see Sect. \ref{sec:avxlf} for details and caveats). Five of them are classified as ultra-luminous infrared galaxies (ULIRGs) and two as luminous infrared galaxies (LIRGs). The remaining object (NGC~4676 or The Mice), is a merging system. 

For these galaxies we verified that the AGN emission does not compromise the IR-base SFR determination. To this end, we used results of  \citet{2003MNRAS.343..585F, 2009MNRAS.399.1373N, 2010ApJ...709..884Y} who used various methods (IR and optical photometry, $5-8~\mu$ Spitzer spectroscopy and SDSS data)  to constrain the contribution of the AGN to the bolometric luminosity. Their results are summarized in the last two columns of the Table \ref{table:sample2}. In all but one galaxy the AGN contribution to the bolometric luminosity is constrained to $\la 30\%$. The remaining galaxy, IRAS~13362+483 (NGC5256), is classified as a Seyfert 2  \citep{2009MNRAS.399.1373N}. However, the $[NV]/[NII]$ line ratios \citep{2011ApJ...730...28P} and absence of the broad line region \citep{2011ApJ...730..121W} in this galaxy suggest that it is starburst-dominated. We therefore decided to keep it in our sample.

Finally, we also included 8 normal galaxies from the Chandra Deep Field North (HDFN) \citep{2001AJ....122.2810B} and Lynx field \citep{2002AJ....123.2223S} compiled by \citet{2003MNRAS.339..793G}. Their selection is described in the above mentioned paper;  their K-corrected  X-ray and radio luminosities  were taken from \citet{2004MNRAS.347L..57G}.  Thus, our high-SFR sample included 16 galaxies, listed in Table \ref{table:sample2} and \ref{table:sample3}.

\subsection{Observer bias}
\label{sec:obs_bias}

Obviously, the selection of galaxies based on the content of the Chandra archive is subject to the observer bias -- we often tend to choose galaxies for our proposals which are  known to be bright in X-rays. This can  shift the scale factor in the obtained  $L_\rmn{X}-\rmn{SFR}$ relation upwards. It may also affect the shape of the luminosity function and that of the $L_\rmn{X}-\rmn{SFR}$ relation. As it is further discussed in the Section \ref{sec:stat}, the behavior of the latter suggests that the observer bias does have some effect on our sample. A straightforward method to assess its amplitude is to compare the $L_\rmn{X}-\rmn{SFR}$ relation obtained here with that  based on galaxy  samples  selected from multi-wavelength surveys. This work, although important, is beyond the scope of this paper and will be addressed in a follow-up publication.

\begin{table*}
\centering
\begin{minipage}{140mm}
\caption{High-SFR sample: LIRGs and ULIRGs.}
\label{table:sample2}
\begin{tabular}{@{}l c c c c c c c c@{}}
\hline
Galaxy & $D$\footnote{distance from \citet{2003AJ....126.1607S}. }  & Luminosity & SFR & M$_{\star}$  & $\rmn{SFR}/M_{\star}$ & $\log L_{\rmn{X}}$ & $\alpha_{\rmn{bol}}$\footnote{AGN contribution to the bolometric luminosity (in per cent).} &Ref.\footnote{References for AGN contribution to the bolometric luminosity:(1)\citet{2009MNRAS.399.1373N}, (2)\citet{2003MNRAS.343..585F}, (3)\citet{2010ApJ...709..884Y}. (4)\citet{2009ApJS..184..230B}} \\
	     & (Mpc)      &	  Class &   ($M_{\odot}$ yr$^{-1}$) & ($10^{10} M_{\odot}$) &  ($10^{-10}$ yr$^{-1}$) & $(\rmn{erg}\,\rmn{s}^{-1})$ & (\%) & \\	
\hline
\hline
IRAS~17208-0014	& 183.0 & ULIRG & 289.9 & 10.3 & 28.1 & 41.40 &$<0.4$ & (1) \\
IRAS~20551-4250	& 179.1 & ULIRG & 139.4 & 7.5  & 18.6 & 41.63 &28$^{+5}_{-4}$ & (1) \\
IRAS~23128-5919	& 184.2 & ULIRG & 139.6 & 7.0  & 19.9 & 41.88 &3.9$^{+1.1}_{-0.8}$ & (1) \\
IRAS~10565+2448	& 182.6 & ULIRG & 156.8 & 9.9  & 15.8 & 41.42 &$<0.1$ & (2) \\
IRAS~13362+4831	& 120.9 & LIRG	& 54.8  & 17.3 & 3.2  & 41.81&Sy 2 & (3) \\
IRAS~09320+6134	& 164.3 & ULIRG & 137.1 & 13.1 & 10.5 & 41.39&$16\pm3$ &(1) \\
IRAS~00344-3349	& 84.0  & LIRG	& 20.1  & 1.5  & 13.7 & 41.23 &$\sim 0$ & (3)  \\
NGC~4676	& 98.2  &     - & 15.8  & 6.6  & 2.4  & 40.92 & $\sim 0$ & (4) \\
\hline
\end{tabular}
\end{minipage}
\end{table*}

\begin{table*}
\centering
\begin{minipage}{140mm}
\caption{High-SFR sample: galaxies from the \textit{Hubble Deep Field} North and Lynx Field.}
\label{table:sample3}
\begin{tabular}{@{}l c c c c c c@{}}
\hline
Galaxy & Redshift & $F_{0.5-8 \rmn{keV}}$ & $L_{0.5-8 \rmn{keV}}$ & $F_{1.4 \rmn{GHz}}$ & $L_{1.4 \rmn{GHz}}$ & SFR \\
 & & ($10^{-15}$ erg s$^{-1}$ cm$^{-2}$)& ($10^{40}$ erg/s) & ( $\umu$Jy) & ($10^{30}$ erg/s) & ($M_{\odot}$ yr$^{-1}$) \\
\hline
123634.5+621213  &  0.456  &   0.43  & 31.43  &  233.00  &   1.63  &  90.56\\
123634.5+621241  &  1.219   &  0.31  &  163.86 & 230.00  &  15.94 &  885.56\\
123649.7+621313  &  0.475   &  0.15   & 10.09  &  49.20  &  0.37  &   20.56\\
123651.1+621030  &  0.410   &  0.30  &  14.61 & 95.00   &  0.52   & 28.89\\
123653.4+621139  &  1.275   &  0.22  &  128.03 &  65.70   &  5.18 &  287.78\\
123708.3+621055 &  0.423   &  0.18  &  9.39  &  45.10  &   0.23  &   12.78\\
123716.3+621512  &  0.232   &  0.18  & 2.55  &  187.00  &   0.27  &   15.0\\
0084857.7+445608   &  0.622   &  1.46  & 177.76 &  320.00  &   4.50 &  250.0\\
\hline
\end{tabular}
Note: See \citet{2003MNRAS.339..793G} for a complete description of the sample; the X-ray and radio luminosities are adopted from \citet{2004MNRAS.347L..57G}.
\medskip
\end{minipage}
\end{table*}

\subsection{A remark regarding the roles of primary and secondary samples}
\label{sec:remark}

It is important to realize the different roles and the difference in the way the two samples are treated. The main goal of this paper is to study population of high-mass X-ray binaries, their luminosity distribution and dependence of their number and collective luminosity on the star-formation rate and other properties of the host galaxy.  This is achieved with the help of resolved galaxies of our primary sample. Their analysis is presently  limited to the emission from compact sources, the contribution of various unresolved emission components is not considered. To properly resolve the populations of compact sources, we considered only  galaxies located within  the distance of $D<40$ Mpc. This limited the volume and, correspondingly the range of SFRs probed by the primary sample. Indeed, the highest SFR in our resolved galaxies does not exceed  $\sim 15-20~M_\odot$/yr. 

In order to probe the high-SFR regime,  we complemented our primary sample with more distant galaxies. This allowed us to extend the dynamical range of SFRs by $\sim$ two orders of magnitude.
On the other hand, due to larger distances to these galaxies, their compact source populations can not be studied at the same level of detail, because of insufficient point source detection sensitivity and source confusion, even in the most nearby of them. Therefore we treated them as unresolved (even though the angular extent of some of them was as large as $\sim 25-50\arcsec$) and only measured their total X-ray emission. Obviously, X-ray luminosities of these galaxies can not be directly compared with our primary sample, as  they include the contribution of diffuse emission, which may be significant in star-burst galaxies.  However these galaxies help to present the broader picture and to constrain the properties of bright HMXBs and ULXs in the high SFR regime and at intermediate redshifts, based on the behavior of their collective emission, using the ideas described in \citet{2004MNRAS.351.1365G}. The quantitative analysis of the full $L_\rmn{X}-SFR$ relation for the two samples, including diffuse X-ray component, will be presented in the forthcoming publication.

\section{X-ray analysis}

\subsection{Data preparation}

We analyzed in total $64$ \textit{Chandra} observations, 63 ACIS-S and one ACIS-I, listed 
in Table \ref{table:tsample}.
The data preparation was done following the standard CIAO\footnote{http://cxc.harvard.edu/ciao3.4/index.html} threads (CIAO version 3.4; CALDB version 3.4.1) and limiting the energy
range to $0.5-8.0$ keV.  
We performed the point source detection in each observation using CIAO \texttt{wavdetect}. The 
scale parameter was changed from the default value. We used the $\sqrt{2}$-series from 1.0 to 
8.0 to have a wide enough range of source sizes to account for the variation in PSF from the inner 
to the outer parts of an observation.
In order to avoid false detections, the value of the parameter \texttt{sighthresh} was kept rather low.
We set it as the inverse of the total number of pixel in the image. The typical values being in the range $\sim 10^{-7}-10^{-8}$.
We used \texttt{maxiter} = 10, \texttt{iterstop} = 0.00001 and \texttt{bkgsigthresh} = 0.0001. 
We set the parameter \texttt{eenergy} = 0.8 (the encircled fraction of source energy 
used for source parameter estimation).

Readout streaks caused by a bright point source were seen in two cases: observation 378 
(M82) and 315 (Antennae). To avoid that such a problem could affect source detection, we 
corrected the events using \texttt{acisreadcorr}.

Multiple observations were present for $\sim 30\%$ of the sample galaxies; they were combined in order to improve the sensitivity and, in case of large objects, to cover the galactic area. We first used CIAO 
\texttt{reproject\_events} to reproject all the observations into the sky coordinates of 
the longest observation of a given galaxy. The images were then merged using \texttt{merge\_all} 
script and the \texttt{wavdetect} task was applied again to the combined image. 
An exposure map weighted according to a specific model for the incident spectrum was computed 
for each observation. 

In computing the exposure maps and counts-to-ergs conversion factor we assume an absorbed power law spectrum with the photon index $\Gamma=2.0$. This choice was motivated by the value of the centroid of the power law photon index distribution of the luminous compact X-ray sources in star-forming galaxies measured by \citet{2004ApJS..154..519S},  $\Gamma=1.97\pm 0.11$. In order to chose the absorption, we produced the combined spectrum of all compact sources in 8 representative galaxies from our sample and determined the best fit value of  $n_{H}\sim 3\times 10^{21}$ cm$^{-2}$. This value used for all galaxies. We note that moderate variations of these parameters do not result in significant changes of the counts-to-ergs conversion coefficient -- change of the slope by $\Delta \Gamma =0.2$ or of the absorption by a factor of 1.5 results in less than $10\%$ change of the conversion factor. This is insignificant for the purpose of our study.

\subsection{Source counts and luminosities}
For X-ray photometry of compact sources we used the approach and scripts from \citet{2007A&A...468...49V}.
The count rate for each detected point source was calculated inside a circular region centered on the source central coordinates given by \texttt{wavdetect} output. 
In order to determine the radius of the circle, for each observation we extracted the point spread function (PSF) using CIAO \texttt{mkpsf} task. In case of multiple observations the PSFs were combined using the values of the exposure maps as weights. We mapped the PSFs into the World Coordinate System (WCS) reference frame of the relative point source image using \texttt{reproject\_image} task. The radius of the circle was determined individually for each source so that the encircled PSF energy was 85\%. For the background region we used an annulus which inner radius was equal to the radius of the source region and the outer radius was 3 times bigger.
The corrected source counts and errors were then found by the following equations  \citep{2007A&A...468...49V}:
\begin{equation}
\label{eq:src_counts}
S=\frac{C(b-d)d^{-1}-Q}{\alpha b d^{-1}-\beta}
\end{equation}
\begin{equation}
\label{eq:src_counts_err}
\sigma^{2}_S=\frac{\sigma^{2}_C(b-d)^{2}d^{-2}+\sigma^{2}_Q}{(\alpha b d^{-1}-\beta)^{2}}
\end{equation}
where $S$ is the number of net counts from the source, $C$ is the number of counts inside the source region and $Q$ is the number of counts in the background region, $\alpha$ is the integral of the PSF over the source region, $\beta$ is the integral of the PSF over the source and background regions, $b$ is the area of the source and background regions and $d$ is the area of the source region. The latter two quantities were derived by integrating the exposure maps rather than computing the geometric areas.

Strictly speaking, the PSF weighting should be based on source counts rather than exposure map values. Indeed, the compact sources under study may show significant variability between observations which may result in the inaccurate determination of the shape of the combined PSF, of the radius of the 85\% region  and, consequently, inaccurate flux calculation. However, several factors limit the practical usefulness of counts-based PSF weighting. Firstly, co-adding observations leads to the increase of the sensitivity, therefore a fraction of sources detected in the summed image are not detected in individual observations. The fraction of such sources is large for the shape of the HMXB luminosity function. Secondly, as the source counts are subject to Poissonian uncertainty, using them as weights in determining the effective PSF will introduce additional statistical uncertainty in the source intensity calculation. For faint sources, this additional uncertainty will exceed the inaccuracy of the exposure map-based procedure. Given the pattern of the {\it Chandra} pointings for the galaxies in our sample, the inaccuracy of the exposure-map-based weighting method is relatively small. Indeed, the galaxies in our sample were the main target of {\it Chandra} observations and therefore were similarly located in the fov of individual observations. With a few exceptions, the analysis regions were located entirely inside a single ccd chip. For these reasons, for a given source, there is not much variation in the size of the $85\%$ circle between individual observations. To verify this, we selected several bright sources located in galaxies with multiple pointings and compared their intensities obtained using the two psf weighting techniques. We found that the difference between the two methods did not exceed $\sim 5-10\%$ (a part of which is due to the additional statistical uncertainty introduced by the counts-based weighting method). For these reasons, for the purpose and parameters of the present work the exposure map -based PSF weighting technique is more appropriate.

The problem of point source crowding may compromise the total X-ray luminosity estimation of galaxies. We identified in each galaxy all the compact sources having background regions overlapping neighboring sources. For the galaxies affected by crowding problem, we performed a second iteration of the circular photometry procedure defining the radius of each source so that it included 90\% of encircled PSF. These regions were then excluded from both the individual image and exposure map of one of the overlapping point sources in order to subtract the source counts and correct the source area respectively. We finally used the corrected image and exposure map to perform the photometry procedure defining the radius of the circle for each source so that the encircled PSF energy was 85\% and obtained the corrected $C$, $Q$, $b$ and $d$ for applying eqs. \ref{eq:src_counts} and \ref{eq:src_counts_err}.

The X-ray luminosities of X-ray binaries in galaxies from our primary sample were computed as a sum of luminosities of compact sources, corrected for the incompleteness and  CXB contribution and transformed to a common luminosity limit, as described in Section \ref{sec:ltot}.

The galaxies from the secondary high-SFR sample were treated as unresolved objects and their luminosity was measured as the background-subtracted luminosity inside the $D25$ ellipse. The counts-to-ergs conversion in this case was determined based on the modeling of their X-ray spectra.
For most of the spectra the best-fit was obtained with a two component model: a thermal plasma plus a power-law corrected for the Galactic absorption. 
In two cases (IRAS~20551-4250 and IRAS~23128-5919) a good fit was obtained with a three-component model, according to \citet{2003MNRAS.343.1181F}: a thermal component and a "leaky -absorber" continuum, including an absorbed plus a non-absorbed power law spectrum of the same photon index.
To measure the X-ray luminosity of unresolved galaxies and the contribution of diffuse gas in resolved galaxies, we created source and background spectra and the associated ARF and 
RMF files using \texttt{specextract} script, which is suitable for extended sources. Spectra were then modeled using XSPEC v. 12.3.1x.

The luminosities of HDFN galaxy sample were measured in $0.5-8\,\rmn{keV}$ band, using their count rates, best fit slopes  from \citet{2001AJ....122.2810B} and redshifts from \citet{2000ApJ...538...29C}.

\begin{table}
\centering
\caption{\textit{Chandra} observations analyzed.}
\label{table:tsample}
\begin{tabular}{l l || l l}
\hline
Galaxy	& Obs-ID 	& \vline\, Galaxy &	Obs-ID \\
\hline
\hline
NGC~0278	&2055	&\vline\, NGC~4631		&797 \\
	      	&2056	&\vline\, NGC~4676	    	&2043 \\		  			       
NGC~0520	&2924	&\vline\, NGC~5194 (M51A)   	&3932 \\
NGC~1313	&2950	&\vline\, 	 	    	&1622 \\			       
NGC~1427A	&3949	&\vline\, NGC~5253		&7153 \\                              
NGC~1569	&782	&\vline\, 		    	&7154 \\  
NGC~1672	&5932	&\vline\, 		    	&2032 \\		  
NGC~2139	&8196	&\vline\, NGC~5457 (M101)   	&934  \\		  
NGC~2403      	&4630	&\vline\, 	 	    	&4732 \\		  
		&4628	&\vline\, 		    	&5309 \\		  
      		&4629	&\vline\, 		    	&5322 \\		  
      		&2014	&\vline\, 		    	&6114 \\
NGC~3034 (M82)	&378 	&\vline\, 		    	&4736 \\		  
      		&2933  	&\vline\, 		    	&4731 \\  
NGC~3079      	&2038  	&\vline\, 		    	&5340 \\	  
NGC~3310	&2939  	&\vline\, 		    	&5300 \\		  
NGC~3556	&2025  	&\vline\, NGC~5474		&9546 \\		  
NGC~3631	&3951  	&\vline\, NGC~5775		&2940 \\		  
NGC~4038/39   	&315   	&\vline\, NGC~7090		&7252 \\
	      	&3040  	&\vline\, 		    	&7060 \\		  
	     	&3041  	&\vline\, NGC~7541		&7070 \\		  
	     	&3042  	&\vline\, NGC~7793		&3954 \\		  
	      	&3043  	&\vline\, IRAS~17208-0014	&2035 \\  
	      	&3044  	&\vline\, IRAS~20551-4250	&2036 \\		  
	      	&3718  	&\vline\, IRAS~23128-5919	&2037 \\	  
NGC~4194	&7071  	&\vline\, IRAS~10565+2448	&3952 \\		  
NGC~4214	&5197  	&\vline\, IRAS~13362+4831	&2044 \\		  
	      	&4743  	&\vline\, IRAS~09320+6134	&2033 \\		  
	      	&2030  	&\vline\, IRAS~00344-3349	&8175 \\	  
NGC~4490      	&4726  	&\vline\, CARTWHEEL		&2019 \\
	      	&4725  	&\vline\, UGC 05720		&9519 \\		  
	      	&1579  	&\vline\, \\ 	  			
NGC~4625	&9549 	&\vline\, \\					
\hline
\end{tabular}
\end{table}

\section{Spatial analysis}
\label{sec:spatial}

\begin{figure}
\begin{center}
\includegraphics[width=1.0\linewidth]{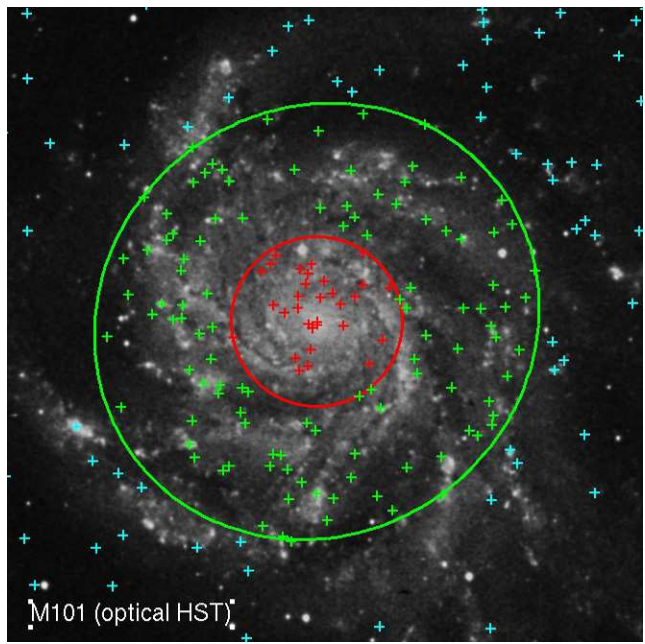}
\caption{Optical (HST) image of M101. The crosses indicate the positions of detected X-ray point sources.  The sources marked in green are used for the further analysis. The part of the galaxy inside the red circle (X-ray sources marked by red crosses) were excluded because their populations were demonstrated to be dominated by LMXBs \citep[see][for details]{2004MNRAS.349..146G}. The region outside green circle (X-ray sources marked by cyan crosses) was excluded because the contribution of CXB sources becomes large. Its size was chosen  as described in Sect.\ref{sec:spatial}.}
\label{fig:selection_example}
\end{center}
\end{figure}

Whereas irregular galaxies host young stars, the stellar population of spirals is spread in a broad range of ages. The old populations of bulges are totally dominated by LMXBs, similar to the ones of elliptical galaxies. An active nucleus can be also present. The disk instead hosts X-ray binaries of different ages. As galaxies are mostly transparent to X-rays above 2 keV, background X-ray 
sources are also detected in each galactic area. Although reliable separation of different classes of sources in external galaxies based on their X-ray properties is virtually impossible, contributions of LMXBs and background AGN can be controlled and subtracted in the statistical manner.
In order to identify regions of galaxies where the HMXBs dominate the compact X-ray source population, we analyzed the spatial distribution of the detected compact sources in resolved galaxies of the primary sample. An example of regions selected based on this method is shown in Fig. \ref{fig:selection_example}.

\subsection{Contribution of the central AGN}
In the resolved galaxies ($D<40$ Mpc), the AGN contribution can be easily discriminated. Three  galaxies from the primary sample are known to have a low luminosity AGN: NGC~1672,  NGC~3079 and  M51A \citep{2006A&A...455..773V}. Whereas the central sources of M51A, NGC~3079 were detected and excluded from the further analysis, no central X-ray source was found in NGC~1672. However,  \citet{2011ApJ...734...33J} found in this galaxy an evidence for a low-luminosity nucleus, with the $2-10$ keV luminosity of $\sim 4\times 10^{38}$ erg s$^{-1}$. Its  non-detection in our analysis is likely a result of different {\tt wavdetect} settings, in particular  of  a higher detection threshold used (in fact, visual inspection of the image did reveal a hump on top of the diffuse emission in the center of the galaxy). 
In addition to these three galaxies we also decided to exclude the compact X-ray source in the center of NGC~4039, although its nature may be somewhat  controversial. \citet{2009ApJ...699.1982B} ruled out the presence
of an AGN in this galaxy, based on the lack of $[Ne V]$ emission and the low $[O IV]/[Ne II]$ ratio.  Moreover, \citet{2009ApJ...690..267D} suggested that the X-ray spectrum is indicative of an X-ray binary in the very high state, rather than of an AGN. However, they pointed out that  location of this source in the nuclear star cluster makes it a likely  LMXB candidate, rather than an HMXB.

For galaxies of the high-SFR sample, the AGN contribution in the X-ray band was constrained directly from the \textit{Chandra} images. Galaxies from the "unresolved" sample of Table \ref{table:sample2} are in fact partly resolved by \textit{Chandra}, the smallest one having angular size of $D25\sim24$ arcsec. This is enough to verify, that none of them has a dominating central point source. The upper limit of 30\% on the AGN contribution to the IR luminosity (see Sect. \ref{sec:sample}) ensures that the IR based SFR determination is not compromised by the AGN. The HDFN and Lynx field galaxies are spiral and irregular galaxies, showing no signs of AGN activity, selected by \citet{2003MNRAS.339..793G} based on the identifications from \citet{1998AJ....116.1039R}.

\subsection{Contribution of LMXBs}
Although only star-forming galaxies with $\rmn{SFR}/M_{\star}\geq 1 \times 10^{-10}$ yr$^{-1}$ have been selected, the contribution of LMXBs to the total X-ray luminosity can be further minimized by excluding bulge sources, which populations are dominated by LMXBs.
This procedure was applied to spiral galaxies having an inclination angle $i \lesssim 65$ deg where the bulge region was identified by inspecting the optical and near-infrared images and taking into account the information available in literature. 
The following spiral galaxies were affected: NGC~0278 ($R_{\rmn{bulge}} = 0.1$ arcmin), NGC~3631 ($R_{\rmn{bulge}} = 0.3$ arcmin), M51~A ($R_{\rmn{bulge}} = 0.4$ arcmin), M101 ($R_{\rmn{bulge}} = 1.5$ arcmin) and NGC~4490 ($R_{\rmn{bulge}} = 0.3$ arcmin). For M101 the bulge region was chosen based on results from \citet{2004MNRAS.349..146G}. The X-ray point sources detected in such regions were then excluded from the analysis. We excluded the same regions from the analysis of infrared and ultraviolet data (Sect. \ref{sec:multiwav}). The fraction of galaxy removed by this procedure ranges between 10\% and 30\%.

\subsection{Contribution of CXB sources}
\label{sec:prof_cxb}
In the central regions of galaxies the surface density of XRBs is sufficiently high that the 
contribution of background AGNs can be neglected. In the outer parts the surface density 
of compact sources becomes comparable to the average density of CXB sources and the latter may become and important source of contamination. In order to minimize this contamination we used the following procedure.

\begin{figure}
\begin{center}
\hbox
{
\includegraphics[width=0.49\linewidth]{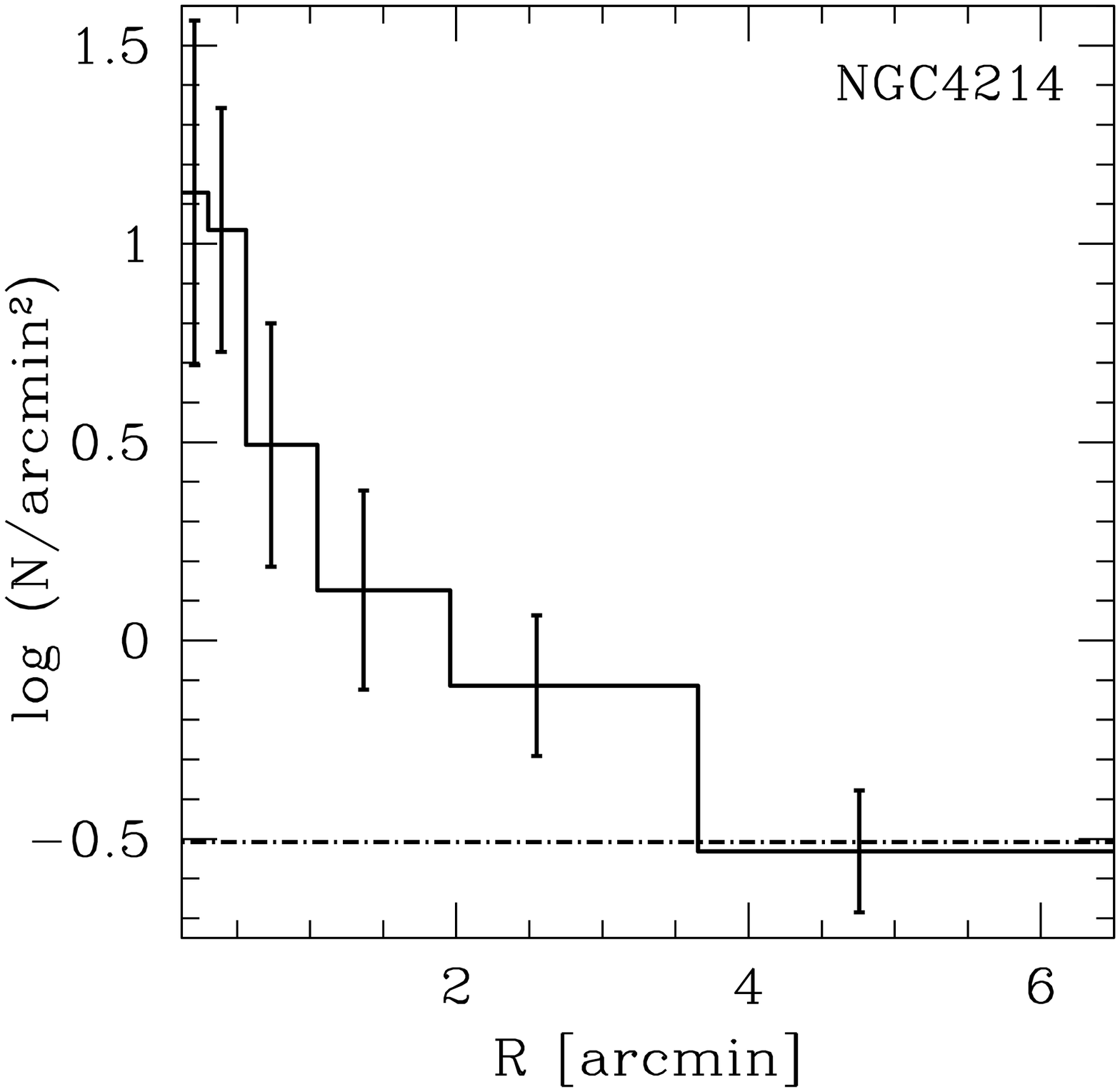}
\includegraphics[width=0.49\linewidth]{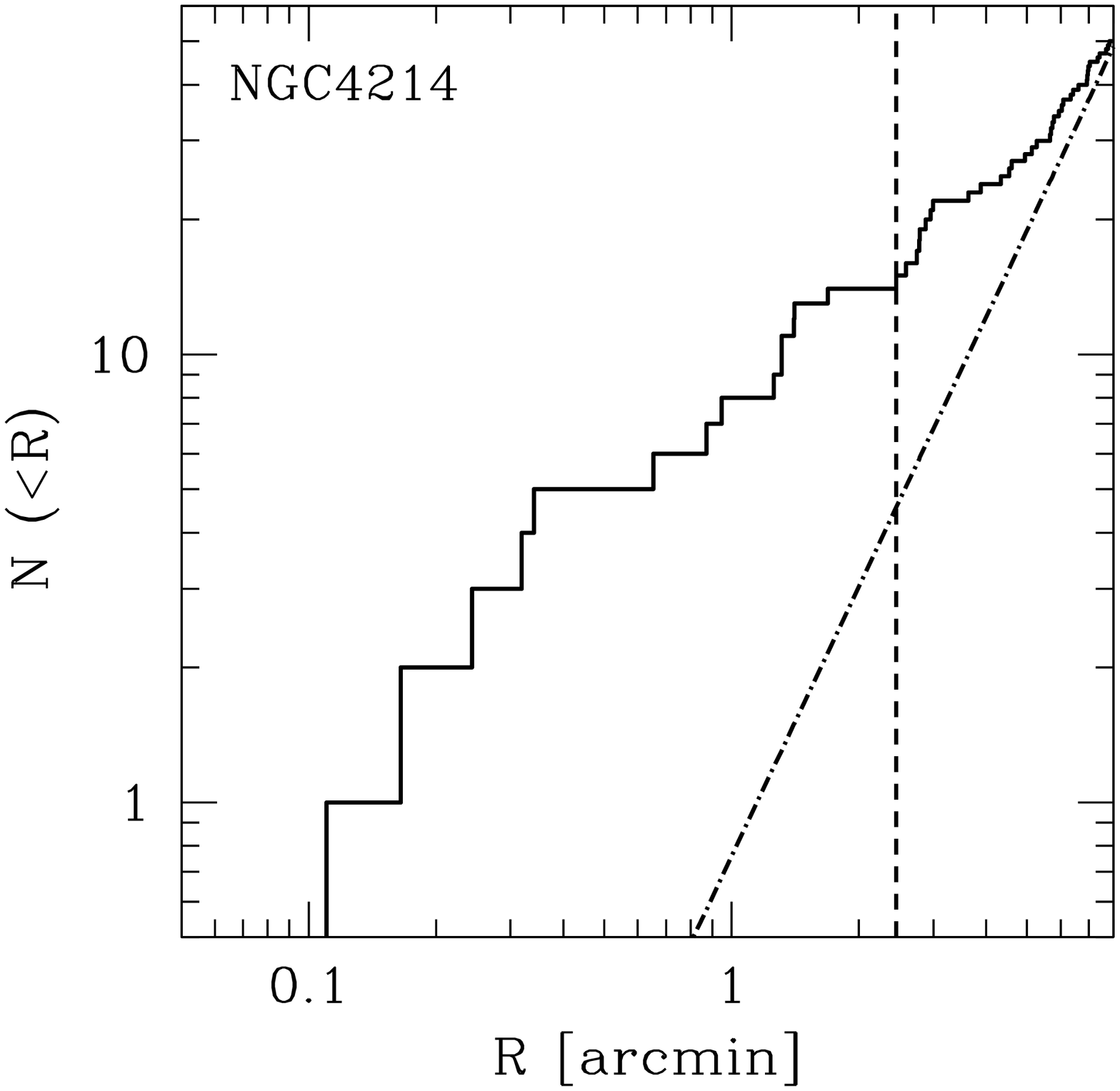}
}
\caption{An example of the spatial analysis performed in resolved galaxies. \textit{Left:} surface density of the detected point sources (solid histogram) plotted together with the predicted CXB level from \citet{2008MNRAS.388.1205G} (dotted-dashed line).  \textit{Right:} cumulative radial profile of detected point sources (solid line) and the CXB growth curve (dotted-dashed line). The vertical dashed line indicates the radius where the cumulative number of CXB sources equals $1/3$ of the total number of detected compact sources. It is the radius  $R_{\rmn{X}}$ of the elliptical region we choose to analyze.}
\label{fig:density_profiles}
\end{center}
\end{figure}

For each resolved galaxy we defined an elliptical region according to the isophotal 
diameter ($D25$), isophotal diameter ratio ($R25$) and the position angle of the major axis ($PA$), taken from the RC3 catalog \citep{1991trcb.book.....D}.
We computed the incompleteness function $K(L)$ of the part of the \textit{Chandra} image within this region, following the method and code from \citet{2006A&A...447...71V}. We defined the completeness luminosity $L_{\rmn{comp}}$ as the luminosity at which $K(L) = 0.8$ (i.e no more than $20\%$ of point sources are missing). For point sources detected with luminosity above the completeness limit, we constructed cumulative and differential radial profiles (see Fig. \ref{fig:density_profiles} for an example) accounting for the elliptical shape of the region. The latter were compared with the CXB level predicted above the same luminosity threshold. We used the full band $\log N-\log S$ of \citet{2008MNRAS.388.1205G} and converted it to the $0.5-8\,\rmn{keV}$ band used for the detection. We checked its consistency with the observed density of point sources using differential profiles (left panel in Fig.\ref{fig:density_profiles}).
The radius $R_{\rmn{X}}$ of the (elliptical) galactic region that will be used for the analysis was defined based on the cumulative profile as the radius within which the contribution of CXB sources equals  $30\%$   of the total number of detected compact sources.
The so determined radii $R_{\rmn{X}}$ approximately follow the relation:
\begin{equation}
R_{\rmn{X}}(\rmn{arcmin})= R\bigg(\frac{N_{\rmn{CXB}}}{N_{\rmn{tot}}}\approx 30\% \bigg) = 1.2\times \bigg(\frac{D25}{2}\bigg)^{0.6}
\end{equation}

In four galaxies (NGC~1427A, NGC~4194, NGC~4625 and UGC 05720) we detected a rather small number ($4-6$) of point sources that did not allow us to perform spatial analysis. For these galaxies we analyzed the $D25$ ellipse regions, as their CXB contamination was found to be in the $5\%-16\%$ range.

\section{Multiwavelength analysis}
\label{sec:multiwav}

For SFR estimations we used data from Spitzer and GALEX archives, the stellar mass was estimated from 2MASS images. For the galaxies from the resolved sample we computed SFR and mass inside the analysis regions defined in the previous section. 
For the galaxies of high-SFR sample we did the measurements in the D25 regions.  

\subsection{Far-infrared}
We analyzed MIPS $70\,\umu$m and $24\,\umu$m Large Field images.
 In particular, we used the "post Basic Calibrated Data" products provided by the Spitzer Space Telescope Data archive\footnote{http://irsa.ipac.caltech.edu/applications/Spitzer/Spitzer/}. These are images calibrated in $\rmn{MJy}/\rmn{sr}$, suitable for photometric measurements.
We did not analyze MIPS $160\,\umu$m images because for most of the galaxies in our sample 
only "Small Field" images were available. This operating mode often covers only a fraction of galaxy. 
For three galaxies (NGC~7541, NGC~2139 and NGC~3631) neither MIPS 
$70\,\umu$m nor $24\,\umu$m data were available. For NGC~3034 (M82) Spitzer data are affected by saturation effects that limit our ability to extract reliable global flux densities.
In these cases we used published \textit{IRAS} data \citep{2003AJ....126.1607S} to compute the IR luminosity.

After having subtracted the background counts, the total net counts were converted from units of 
$\rmn{MJy}/\rmn{sr}$ into $\rmn{Jy}$ using the following conversion factors, which take into 
account the pixel size of MIPS $70\,\umu\rmn{m}$ and $24\,\umu\rmn{m}$ detectors:
\begin{equation}
\label{eq:conv7024}
\begin{array}{c c}
C_{70\,\umu\rmn{m}}=3.76\times10^{-4}; & C_{24\,\umu\rmn{m}}=1.41\times10^{-4} \\
\end{array} 
\end{equation}
The monochromatic fluxes (Jy) at $24\,\umu$m and $70\,\umu$m were then converted into spectral luminosities ($\rmn{erg}/\rmn{s}$/Hz). The total IR luminosity ($8-1000\,\umu$m) was estimated using the relations from \citet{2008A&A...479...83B}, which give $L_{IR}$  in units of solar luminosity  $L_{\odot} = 3.839\times 10^{33}$ erg/s.:
\begin{equation}
\label{eq:lir_l70}
L_{\rmn{IR}} (L_{\odot}) =7.90\times (\nu L_{\nu})_{70\,\umu\rmn{m, rest}}^{0.94}
\end{equation}
\begin{equation}
\label{eq:lir_l24}
L_{\rmn{IR}} (L_{\odot}) =6856\times (\nu L_{\nu})_{24\,\umu\rmn{m, rest}}^{0.71}
\end{equation}
We preferentially used equation (\ref{eq:lir_l70}), as it has a $\sim 3$ times smaller dispersion than eq. (\ref{eq:lir_l24}) \citep[see][for details]{2008A&A...479...83B}.

\subsection{Near-infrared}
The near-infrared analysis was performed using the $K_{\rmn{S}}$ (2.16$\,\umu$m) images and mosaics 
of 2MASS All-Sky Extended Source Catalog\footnote{http://irsa.ipac.caltech.edu/applications/2MASS/PubGalPS/} (XSC) and Large Galaxy Atlas\footnote{http://irsa.ipac.caltech.edu/applications/2MASS/LGA/} (LGA). 
We converted the source counts measured on images into calibrated magnitudes using 
the zero point magnitude KMAGZP given in the image header:
\begin{equation}
\label{eq:m_k}
m_{\rmn{K}}(\rmn{mag}) = \rmn{KMAGZP} - 2.5 \log (S)
\end{equation}
where $S$ is the integrated, background-subtracted flux in data-number units ("DN") measured within the selected galactic region. After having obtained $K_{\rmn{S}}$ magnitudes, we estimated the $K_{\rmn{S}}$-band luminosity and, following \citet{2004MNRAS.349..146G}, the stellar mass.

\subsection{Ultraviolet}
\label{sec:uv}

We based the ultraviolet analysis on GALEX NUV and FUV background-subtracted intensity maps publicly available in the archive of GR4/GR5 Data Release\footnote{http://galex.stsci.edu/GR4/?page=mastform}. These are images calibrated in units of counts 
per pixel per second, corrected for the relative response, with the background removed. 
We converted the net counts to flux using the conversion factors\footnote{http://galexgi.gsfc.nasa.gov/docs/galex/FAQ/counts\_background.html} between GALEX count rate ($\rmn{ct}/\rmn{s}$) and flux (erg/cm$^{2}$/s/$\AA$):
\begin{equation}
\label{eq:conv_fuv_nuv}
\begin{array}{c c}
 C_{\rmn{FUV}}=1.40\times 10^{-15}; & C_{\rmn{NUV}}=2.06\times 10^{-16} \\
\end{array} 
\end{equation}
The fluxes were then converted into monochromatic luminosities ($\rmn{erg}/\rmn{s}/\AA$) at 
$1529\,\AA$ (FUV) and $2312\,\AA$ (NUV). For three galaxies of our sample (NGC~0278, NGC~4194 and NGC~7541) neither NUV nor FUV data were available. Their treatment is discussed later.

\begin{figure}
\begin{center}
\includegraphics[width=1.0\linewidth]{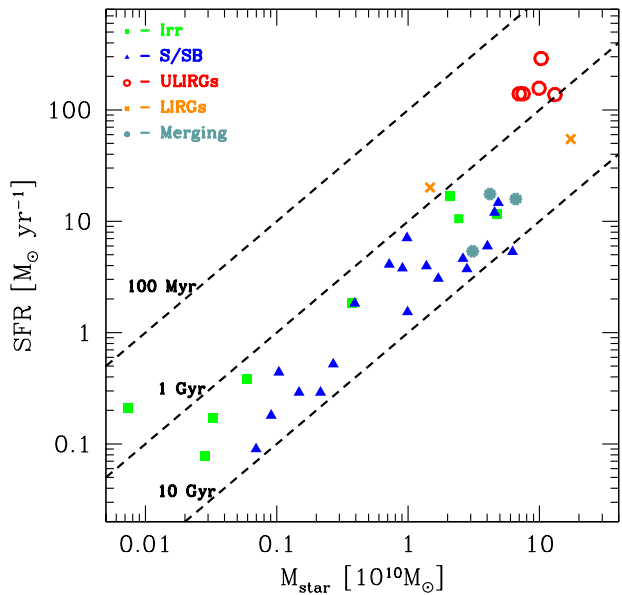}
\caption{The $\rmn{SFR}-M_{\star}$ plane.  Different types of galaxies are plotted with different symbols. The dashed lines correspond to constant stellar-mass-to-SFR ratio.}
\label{fig:age_lines}
\end{center}
\end{figure}

\section{Star formation rate}
\label{sec:sfr}

We used a complex SFR proxy  that takes into account both the UV light escaping the galaxy and the IR emission of the dust heated by young stars \citep{2003ApJ...586..794B, 2003A&A...410...83H, 2004A&A...419..109I, 2006ApJS..164...38I}. 
It has the advantage of being free of the model dependence for the attenuation correction of the UV emission. It contains also a correction factor, $\eta$, accounting for the fraction of the IR emission due to old stellar population in normal star-forming galaxies: 
\begin{equation}
\label{eq:sfr_tot}
 \rmn{SFR_{tot}} = \rmn{SFR_{NUV}^{0}}+(1-\eta) \rmn{SFR_{IR}}
\end{equation}
where $\rmn{SFR_{NUV}^{0}}$ and $\rmn{SFR_{IR}}$ are obtained assuming a Salpeter 
IMF from 0.1 to 100 $M_{\odot}$:
\begin{equation}
\label{eq:sfr_ir}
 \rmn{SFR_{IR}} (M_{\odot}\,\rmn{yr}^{-1}) = 4.6 \times 10^{-44} L_{\rmn{IR}} (\rmn{erg}\,\rmn{s}^{-1})
\end{equation}
\begin{equation}
\label{eq:sfr_nuv_obs}
\rmn{SFR_{NUV}^{0}} (M_{\odot}\,\rmn{yr}^{-1}) = 1.2 \times 10^{-43} L_{\rmn{NUV,obs}} (\rmn{erg}\,\rmn{s}^{-1})
\end{equation}
$L_{\rmn{ NUV,obs}}$ is the observed NUV ($2312\,\AA$) luminosity, i.e. uncorrected for dust 
attenuation, and $L_{\rmn{IR}}$ is the $8-1000\,\umu$m luminosity. 
Eqs. (\ref{eq:sfr_ir}) and (\ref{eq:sfr_nuv_obs}) by \citet{2006ApJS..164...38I} are consistent with those of 
\citet{1998ARA&A..36..189K} within $3\%$ and $10\%$ respectively.
The second term in eq. (\ref{eq:sfr_tot}) is very similar to the calibration of 
\citet{2003ApJ...586..794B} (their eq. (5)).
The quantity $\eta$ is the cirrus correction, i.e. the fraction of IR luminosity due to old stellar population.  It depends on the type of the galaxy but accurate values for individual galaxies are difficult to determine, therefore average values are usually used. We use results of \citet{2003A&A...410...83H}:
\begin{equation}
\eta \approx \begin{cases}
0.4 & \quad \mbox{for normal disk galaxies}\\
0 &\quad \mbox{for starbursts}\\
\end{cases}
\end{equation}
To be consistent with \citet{2003A&A...410...83H}, we used the atlas of \citet{1993ApJS...86....5K} 
to classify the objects of our sample as starbursts or normal star-forming galaxies and use the 
appropriate value of $\eta$ in computing the $\rmn{SFR}$.
This definition is similar, but not identical to the one used by \citet{2003ApJ...586..794B} ($\eta \sim 0.32$ for 
galaxies having $L_{\rmn{IR}} > 10^{11} L_{\odot}$ and $\eta \sim 0.09$ for galaxies having 
$L_{\rmn{IR}} \leq 10^{11} L_{\odot}$).

For each of the three galaxies with no UV data available (see Sect. \ref{sec:uv}), we determined the average value of the  $L_{\rmn{ NUV,obs}}/L_{\rmn{IR}}$ ratio for galaxies of similar Hubble type and inclination. We obtained values in the range $0.08-0.4$. This ratio was used to estimate $L_{\rmn{ NUV,obs}}$ from $L_{\rmn{IR}}$ and then eq.(\ref{eq:sfr_tot}) was applied.

The SFRs for HDFN galaxies were computed based on their 1.4 GHz luminosities and calibration of \citet{2003ApJ...586..794B}:
\begin{equation}
\label{eq:sfr_radio}
 \rmn{SFR} (M_{\odot}\,\rmn{ yr}^{-1}) = 5.55\times 10^{-29}L_{1.4\,\rmn{GHz}}(\rmn{erg}\,\rmn{s}^{-1})
\end{equation}
In Fig. \ref{fig:age_lines} we present the two samples of star-forming galaxies in the $\rmn{SFR}-M_{\star}$ plane. 

\begin{figure}
\includegraphics[width=1.0\linewidth]{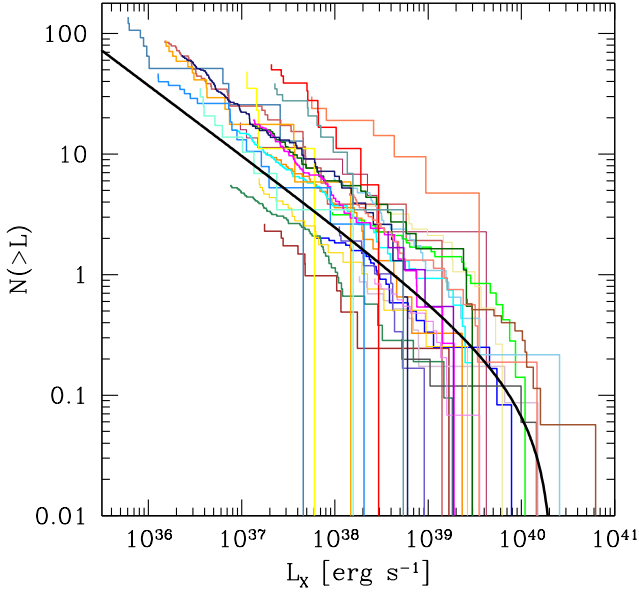}
\caption{Cumulative X-ray luminosity functions of the galaxies from primary sample, normalized by their respective SFRs. The solid line is the cumulative XLF per unit SFR,  given by integration of the equation (\ref{eq:dlf}).}
\label{fig:cxlf}
\end{figure}

\begin{figure*}
\begin{center}
\hbox
{
\includegraphics[width=0.5\linewidth]{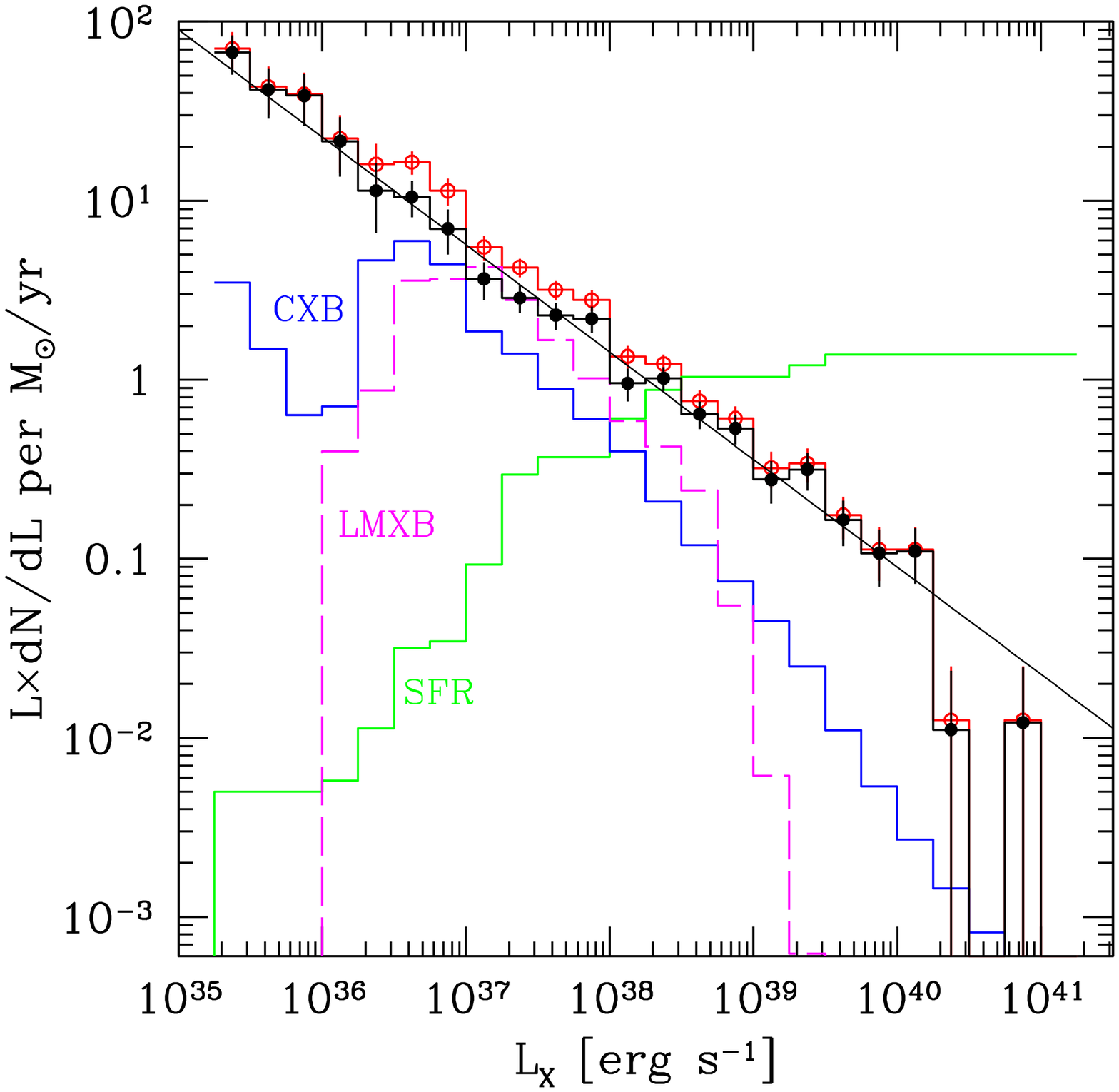}
\includegraphics[width=0.5\linewidth]{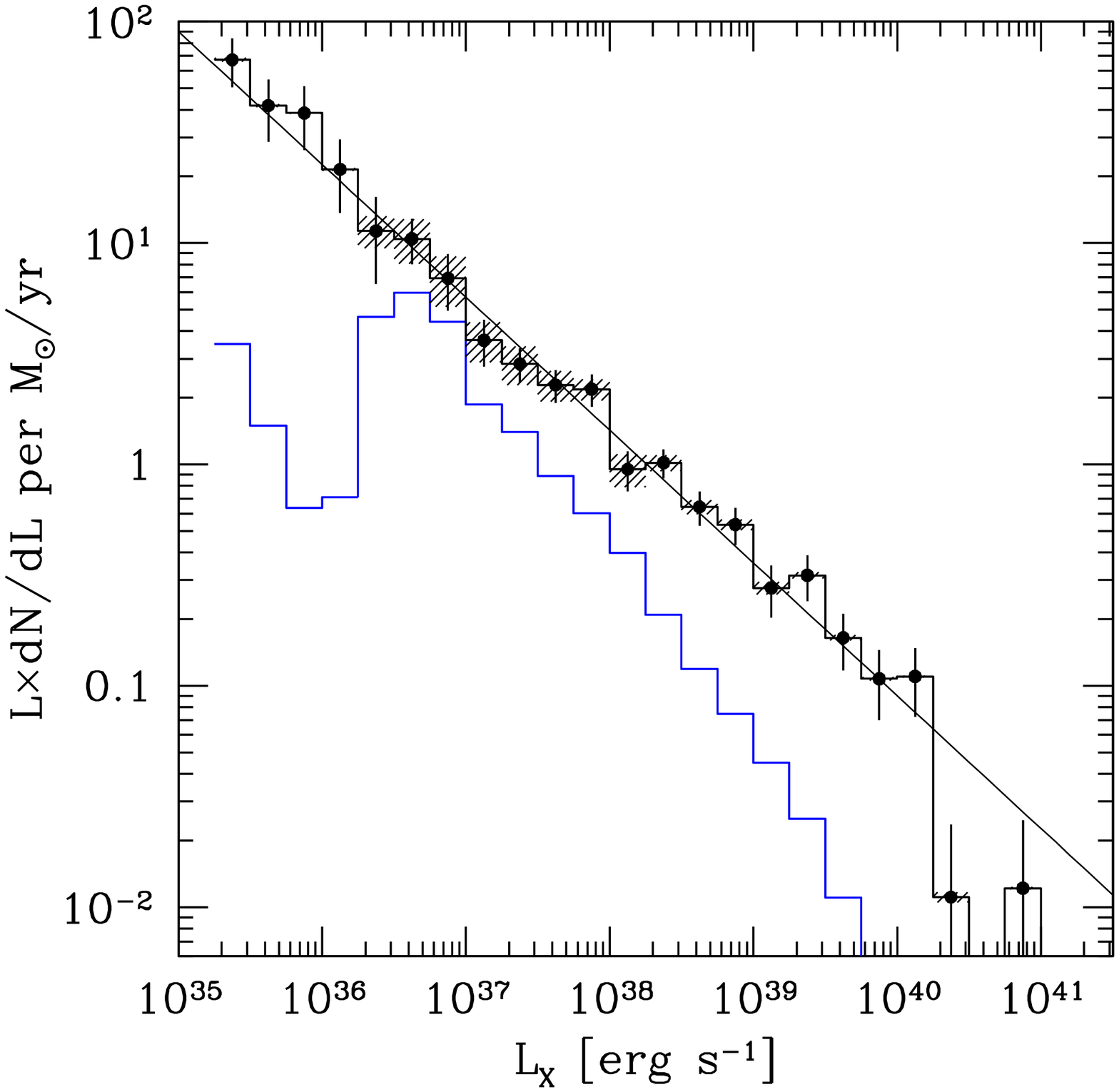}
}
\caption{The luminosity functions obtained by combining data of all the galaxies of the primary sample, normalized by SFR.  \textit{Left:} The top histogram with error bars (open circles, red) shows the luminosity distribution of all compact sources. The two histograms peaked near $\log(L_X)\sim 37$ are predicted contributions of resolved CXB (blue) and LMXB (magenta, dashed) sources. The latter was computed using the average scaling relations for nearby early type galaxies from \citet{2004MNRAS.349..146G}, irrespective of their age. Apparently, it  grossly overestimates the LMXB numbers, therefore no attempt to subtracts their contribution was made. The histogram marked with solid circles (black) shows the luminosity distribution of compact sources with the CXB contribution subtracted. It is our best approximation to the average  HMXB XLF in nearby galaxies. The histogram marked "SFR" (green) shows the total SFR (in unit of $100\, M_{\odot}\,\rmn{ yr}^{-1}$) of all galaxies contributing to a given luminosity bin in the XLF.   \textit{Right:} The CXB subtracted luminosity distribution from the left panel. The shaded area indicates the range of uncertainties corresponding to 40\% uncertainty in the CXB normalization. The solid line on both panels shows a power law with parameters according to eq.(\ref{eq:dlf}).}
\label{fig:dxlf}
\end{center}
\end{figure*}

\section{The luminosity function of HMXB}
\label{sec:xlf}

The  Fig. \ref{fig:cxlf} shows  cumulative luminosity distributions for all galaxies from the primary sample, normalized to their respective SFRs.  It is apparent that although shapes of the distributions are similar, there is a considerable dispersion in their normalization. Their amplitude and significance will be discussed later in this section.

\subsection{Average XLF of HMXBs}
\label{sec:avxlf}

We construct the average luminosity function combining the data for all resolved galaxies. It includes over 700 compact sources. An intuitive and straightforward method would be to bin the  sources into e.g. logarithmically spaced bins and normalized the result by the sum of the SFR of all galaxies contributing to the given luminosity bins. The latter step is required to account for the fact that different galaxies have different point source detection sensitivity.  However, because of the considerable dispersion in the normalization (Fig.\ref{fig:cxlf}), the so computed luminosity function may have a number of  artificial steps and features  at the luminosities, corresponding to the luminosity limits of particular galaxies. In order to deal with this problem we used the following method.

Considering that there is much larger dispersion in normalization than in the shape of the XLF in individual galaxies, we write for the luminosity distribution in the k-th galaxy:
\begin{equation}
\left(\frac{dN}{dL}\right)_k=\xi_k\, \rmn{SFR}_k\, f(L)
\end{equation}
where $SFR_k$ is star-formation rate in the k-th galaxy, $\xi_k$ -- the XLF normalization and the function $f(L)$ describes the XLF shape, assumed to be same in all galaxies, which we would like to determine. 
The number of sources in the j-th luminosity bin, $\Delta N(L_j)$, is:
\begin{equation}
\Delta N(L_j)=\sum_k \Delta N_k(L_j)=f(L_j)\, \Delta L_j \, \sum_k \rmn{SFR}_k \xi_k
\end{equation}
where $\Delta N_k(L_j)$ is the CXB-subtracted number of compact sources in the k-th galaxy in the j-th luminosity bin and  summation is performed over all galaxies of the sample.
The $f(L)$ can be determined as:
\begin{equation}
f(L_j)=\frac{1}{ \Delta L_j }\, \frac{\sum_k \Delta N_k}{\sum_k \rmn{SFR}_k\, \xi_k}
\label{eq:fl}
\end{equation}

For a power law luminosity function $f(L)=L^{-\gamma}$ the $\xi_k$ are calculated as:
\begin{equation}
\xi_{k} = \frac{1-\gamma}{\rmn{SFR}_k}\, \frac{N_{k}(>L_{\rmn{th},k})-N_{\rmn{CXB}}(>F_{\rmn{th},k})}{(L_{\rmn{th},k}^{-\gamma+1}-L_{\rmn{cut}}^{-\gamma+1})}
\label{eq:ksik}
\end{equation}
where $L_{\rmn{th},k}$ is the sensitivity limit for the k-th galaxy,   $N_{k}(>L_{\rmn{th},k})$ is the number of sources detected above this sensitivity limit,  $N_{\rmn{CXB}}(>F_{th,k})$ is the predicted number of resolved CXB sources above the corresponding flux limit  $F_{th,k}=L_{th,k}/4\pi D_k^2$,  $L_{cut}$ is the high luminosity cut-off of the XLF. As it is obvious from eq.(\ref{eq:fl}) and (\ref{eq:ksik}), the so computed $f(L)$ is independent from $\rmn{SFR}_k$.

To compute $\xi_k$ we used  a two-step iterative procedure. At the first step we assumed $\gamma=1.6$, $L_{\rmn{cut}}=2.1\cdot 10^{40}$ erg/s \citep{2003MNRAS.339..793G} and computed the XLF using eqs.(\ref{eq:fl}) and (\ref{eq:ksik}).
The $L_{\rmn{th},k}$ was set to 1.5 times the completeness limit for each galaxy. The obtained luminosity function was fit with a broken power law
\begin{equation}
\label{eq:dlf}
\frac{\rmn{d}N}{\rmn{d}L_{38}} = \xi\,\rmn{SFR}\times 
\begin{cases}
L_{38}^{-\gamma_{1}}, & \quad L_{38}<L_{b}\\
&\\
L_{b}^{\gamma_{2}-\gamma_{1}}\,L_{38}^{-\gamma_{2}}, & \quad L_{b} \leq L_{38}\leq L_{\rmn{cut}}\\ 
\end{cases}
\end{equation}
where $L_{38}=L_{\rmn{X}}/10^{38} \rmn{erg}/\rmn{s}$, $L_{b}$ is the break luminosity, $\rmn{SFR}$ is in units of $M_{\odot}\, \rmn{yr}^{-1}$ and $\xi$ is the average normalization. 
The fit was done on unbinned data using maximum-likelihood (ML) method, the predicted contribution of resolved CXB sources was included in the model as described below. Using this model and its best-fit parameters we calculated more accurate values of $\xi_{k}$ and repeated the procedure. 
We obtained the following best-fit values for the XLF parameters: 
$\gamma_{1}=1.58\pm 0.02$, $\gamma_{2}=2.73^{+1.58}_{-0.54}$, $L_{b}=110^{+57}_{-34}$, $\xi = 1.49 \pm 0.07$. The value of the high-luminosity cut-off was fixed at $L_{\rmn{cut}} = 5\times 10^{3}$, the result being virtually independent on this choice. The final luminosity distribution is shown in Fig. \ref{fig:dxlf}. The obtained XLF and its parameters are remarkably similar to the one determined by \citet{2003MNRAS.339..793G}. Correspondingly, the second step of this procedure does not has any noticeable effect. 

The statistical uncertainty of the slope $\gamma_{1}$ of the average XLF, quoted above is rather small, thanks to the large number of sources involved in its determination. The rms of individual slopes around their average value is much bigger, and is investigated in the section \ref{sec:xlfs}

\subsection{High-luminosity break}

The best fit XLF model has a break at $L_{\rmn{X}}\approx 1.1\times 10^{40}$ erg/s. To access its significance we compare observed number of bright sources with the predictions of the single slope power law model. The best fit parameters of the latter are: $\gamma=1.60\pm 0.02$, $\xi = 1.49 \pm 0.07$. This model predicts $\approx 21.1$ sources above $L_\rmn{X}=10^{40}$ erg/s, whereas we detected 11 sources. The discrepancy between the data and the model is $\approx 2.5\sigma$, without account for the arbitrariness in the choice of the threshold luminosity, i.e. is marginally significant.

\begin{figure}
\begin{center}
\vspace{0.3cm}
\includegraphics[width=0.95\linewidth]{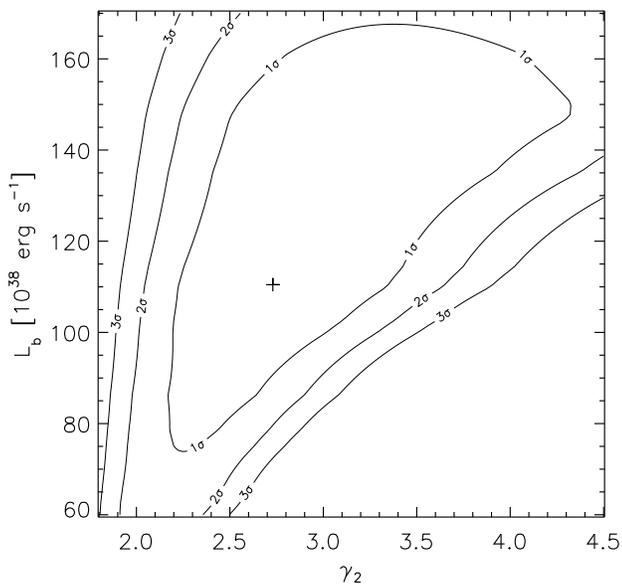}
\caption{The error contours for the XLF break parameters:  break luminosity $L_b$ and the power law index after the break $\gamma_2$. The cross indicates the best fit values.}
\label{fig:contours}
\end{center}
\end{figure}

The error contours for the break luminosity and the XLF slope after the break are shown in Fig. \ref{fig:contours}, demonstrating that the break parameters are not very well constrained, which is not surprising given its low statistical significance.

\begin{figure}
\begin{center}
\includegraphics[width=1\linewidth]{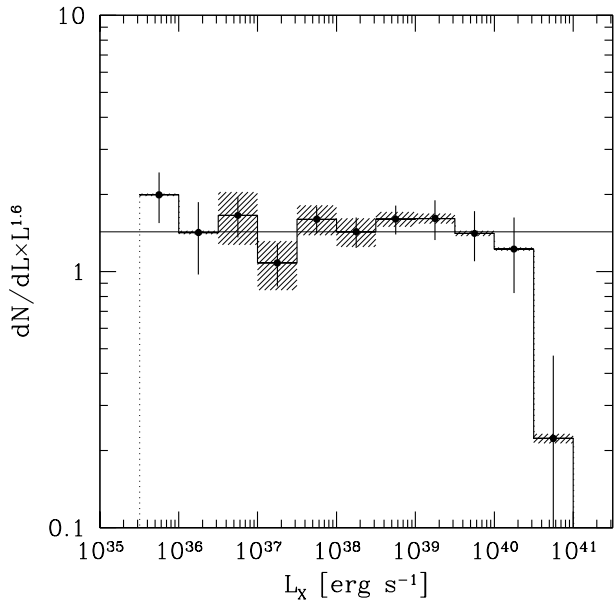}
\caption{The CXB subtracted XLF of compact sources, multiplied by $L^{1.6}$. The shaded are has same meaning as in Fig.\ref{fig:dxlf}.}
\label{fig:dxlf2}
\end{center}
\end{figure}

\subsection{Contribution of LMXBs and CXB sources}
\label{sec:lmxb_cxb}

Expected contributions of resolved CXB and LMXB sources to the XLF  are shown in  Fig. \ref{fig:dxlf}. They were computed summing up individual predictions for all galaxies contributing sources to the given luminosity bin and then renormalizing the total in the same way as the XLF itself. For LMXBs we used the average scaling relation for nearby early type galaxies  from  \citet{2004MNRAS.349..146G}, for CXB sources we used  the $\log N-\log S$ determined by \citet{2008MNRAS.388.1205G}. 

Both predictions bear uncertainties which may affect the resulting HMXB XLF. In the case of the CXB contribution, the uncertainties are caused by angular fluctuations of the density of background AGN and are likely to be reduced as a result of averaging over a rather large number of galaxies involved in the calculation of XLF. However, their final amplitude  is difficult to estimate, we therefore illustrate their possible effect  by showing the range of XLF variations corresponding to 40\% variations in the CXB normalization (shaded area in the right-hand panel in Fig.\ref{fig:dxlf}). As clear from the Fig.\ref{fig:dxlf}, the CXB contribution is significant and its uncertainties are relatively important in the $\log(L)\sim 36-38.5$    luminosity range. It is also clear from the figure that the CXB contribution is predicted with a reasonable accuracy -- the amplitude of the feature in the total XLF at $\log(L)\sim 36-37$ is consistent  with the expected CXB contribution. 

Estimation of the LMXB  contribution is less straightforward.
There is an evidence that the LMXB population may have a rather strong age dependence, being smaller for younger galaxies \citep{2010A&A...512A..16B}. As the scaling relation of  \citet{2004MNRAS.349..146G} was computed for a sample of nearby early type galaxies irrespective of their age, it may be not directly  applicable to galaxies with significant ongoing star-formation.   Indeed, the calculation based  on results of  \citet{2004MNRAS.349..146G} apparently over-predicts LMXB contribution, probably by the factor of a few -- its subtraction from the XLF would result in negative values  (Fig.\ref{fig:dxlf}). For this reason we decided  to subtract only the contribution of CXB sources, until detailed information on scaling relations for LMXBs in late type galaxies becomes available. The part of the XLF affected by this uncertainty is in the $\log(L_\rmn{X})\sim 36.5-38.5$ luminosity range. 

We note that the bright end of the XLF is virtually free of both contaminating components -- the vast majority of sources brighter than $\sim 10^{39}$ erg/s in our sample are high mass X-ray binaries. 

\begin{figure*}
\begin{center}
\hbox
{
\includegraphics[width=0.5\linewidth]{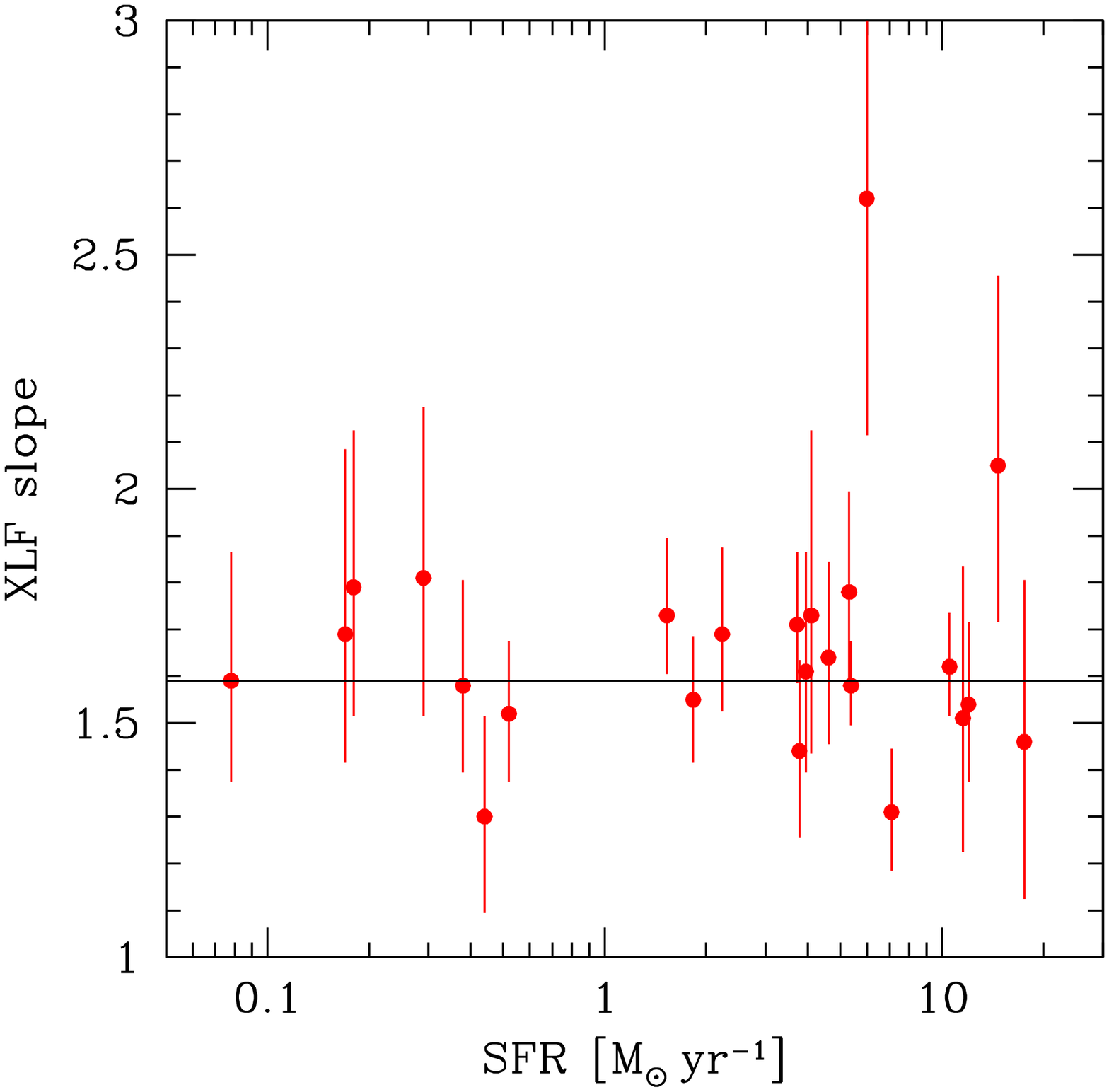}
\includegraphics[width=0.5\linewidth]{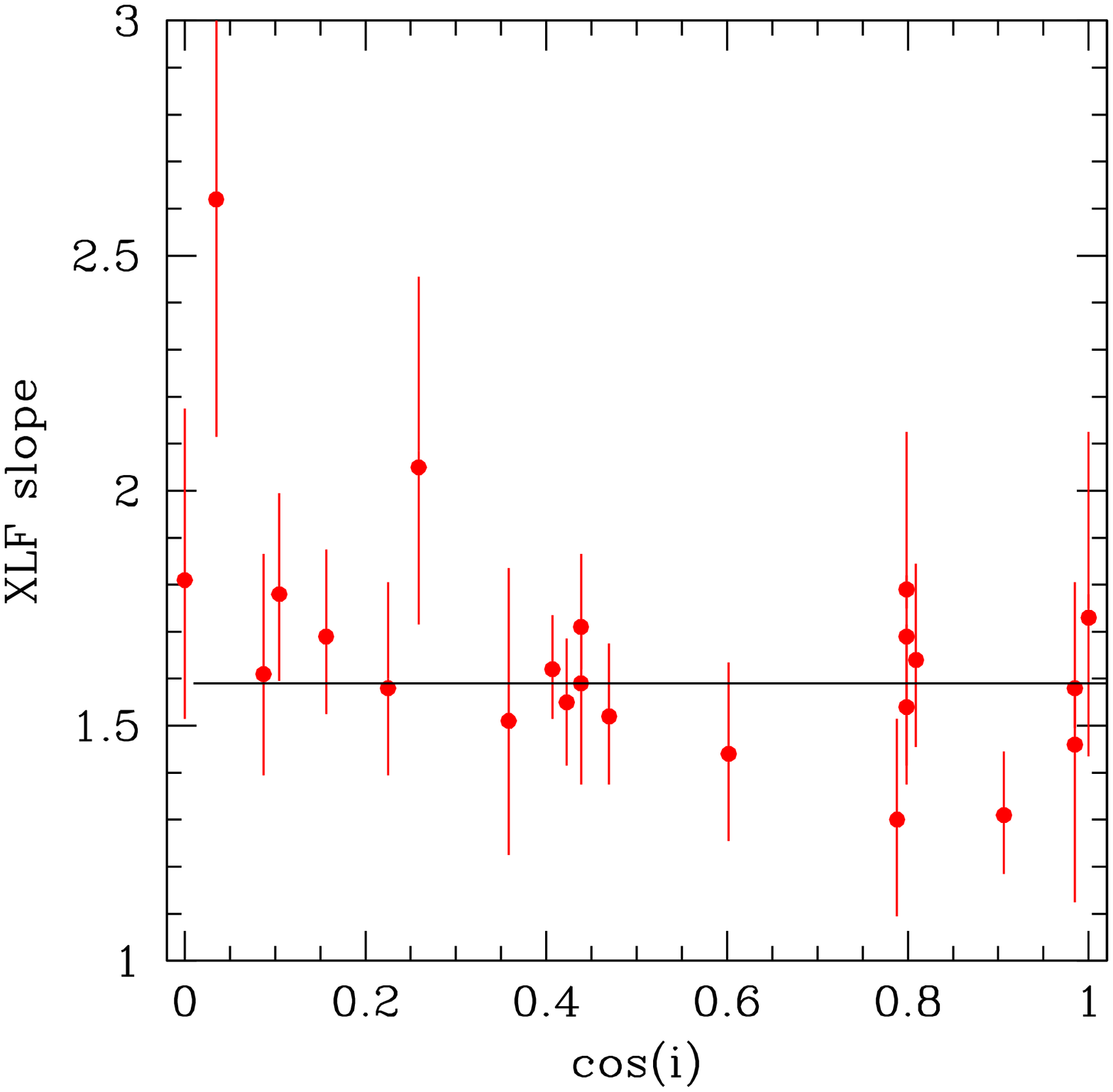}
}
\caption{\textit{Left:} XLF slopes for individual galaxies  plotted against the SFR (left) and inclination (right) of the galaxy. The horizontal solid line shows the mean value of $\left<\gamma\right>\approx 1.6$.}
\label{fig:slopes}
\end{center}
\end{figure*}

\subsection{XLFs of individual galaxies}
\label{sec:xlfs}

We fit XLFs of individual galaxies with a single slope power law with a high luminosity cut-off
\begin{equation}
\frac{dN}{dL_{38}}=A\times L_{38}^{-\gamma}, ~~~~~L\le L_\rmn{cut}=10^{41} {\rm ~erg/s}
\label{eq:xlf1}
\end{equation} 
The cut-off luminosity was fixed at the value, exceeding the luminosity of the brightest source in the sample.  Only galaxies  having more than 5 sources above the detection threshold were analyzed,  the detection threshold being defined as 1.5 times  the 80\% completeness limit of a given galaxy. The CXB contribution, computed according to  \citet{2008MNRAS.388.1205G}, was included as a fixed component of the model. The fit was done with the unbinned data, using the Maximum Likelihood method.

The best-fit values of slope are shown in Fig.\ref{fig:slopes}. The average (computed using $\chi^2$ method) is $\left<\gamma\right>=1.59$ with $rms=0.25$. The $\chi^2=16.8$ for 24 degrees of freedom indicates that the data is consistent with the $\gamma$ being same in all galaxies. The two galaxies deviating from the mean by $\sim 2\sigma$ are NGC~3079 ($\gamma=2.62^{+0.64}_{-0.50}$) and NGC~3310 ($\gamma=1.30^{+0.22}_{-0.20}$). 

There is some evidence that the XLF slope depends on the inclination of the galaxy (right panel in Fig.\ref{fig:slopes}), but it is not statistically significant. The $\chi^2=14.2$ for a linear model $\gamma=\gamma_0+k\times \cos(i)$ with the best fit parameters of $\gamma_0=1.71\pm 0.09$ and $k=-0.22\pm 0.14$. According to the F-ratio test such a reduction of $\chi^2$ corresponds to a $\approx 2\sigma$ fluctuation.  We note however, that the character of the correlation (steeper XLF slopes for more edge-on galaxies) is in accord with the theoretical expectation. Indeed, as the brighter sources (black holes in the soft spectral state) tend to have softer spectra, they are more subject to absorption which is stronger in more edge-on galaxies. This makes the bright end of the XLF to be suppressed in more edge-on galaxies, i.e. their XLFs are expected  to be steeper, as observed.
No other physically meaningful correlations were seen in the data.

\begin{figure}
\begin{center}
\includegraphics[width=1.0\linewidth]{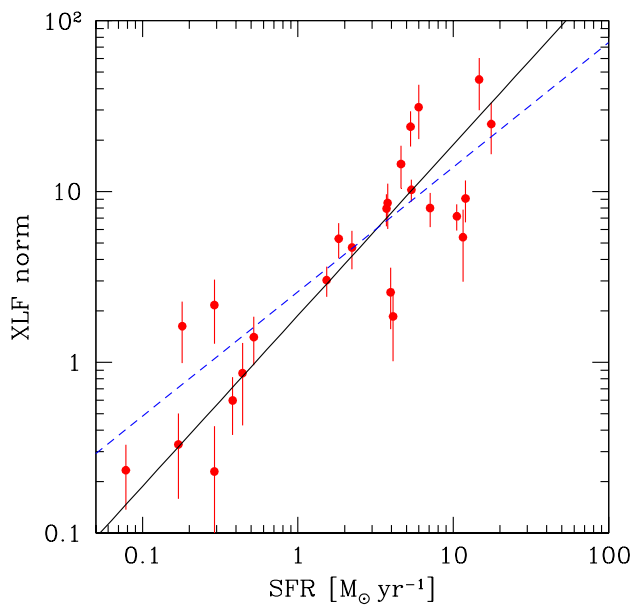}
\caption{The XLF normalization versus SFR. The solid line shows the best-fit linear relation, dashed line - the best fit relation in the form $A=\xi\times \rmn{SFR}^{\beta}$. See text for details.} 
\label{fig:xlfnorm}
\end{center}
\end{figure}

Dependence of the XLF normalization $A$ on the SFR is shown in Fig.\ref{fig:xlfnorm}. The $\chi^2$ fit to the logarithm of the normalization in the form $\log A=a+\beta\log \rmn{SFR}$ yields $\beta=0.73\pm 0.05$ with very large value of the $\chi^2=99.2$ for 23 d.o.f and the $rms=0.32$ dex. 
Forcing a linear relation  $A=\xi\times \rmn{SFR}$ we obtained $\xi=1.88$ with significantly larger $\chi^2=130.4$ for 24 d.o.f., but with $rms=0.34$ dex,  just marginally worse than obtained in the non-linear fit. Both best-fit models are shown in Fig.\ref{fig:xlfnorm}.

The relation between $A$ and $\rmn{SFR}$ can be expressed in terms of the relation between the number of sources above a certain luminosity limit and the star-formation rate:
\begin{equation}
\label{eq:N_sfr}
N_{\rmn{XRB}}(L>10^{38} \rmn{erg}\,\rmn{s}^{-1}) = 3.22 \times \rmn{SFR}\,(M_{\odot}\, \rmn{yr}^{-1})
\end{equation}
For our sample of galaxies this relation predicts the number of sources with the accuracy of $rms=0.34$ dex i.e. within a factor of $\sim 2$ (rms).

The $\beta\ne 1$ obtained above implies a possible non-linear slope of the relation between XLF normalization (i.e. the number of sources) and SFR. However, neither linear nor non-linear model describes the data adequately ($\chi^2/d.o.f.\sim 4-5$), suggesting that additional parameter(s) play a significant  role in this relation. Therefore it is not clear how much weight should be given to the non-linear best fit slope of the $N_{\rmn{X}}-\rmn{SFR}$ relation obtained above. It may, for example, be a result of application of  an inadequate or incomplete  model to a  particular selection of galaxies. Alternatively, it may point at a genuine non-linear relation between the number of sources and SFR. A clue may be provided by accurate characterization of the factors which can potentially affect the number of HMXBs in star-forming galaxies, such as metallicity and/or recent star-formation history. With the presently available data, it does not seems to be possible to discriminate between these two possibilities.  
We note  that \citet{2003MNRAS.339..793G}, based on a much smaller sample of galaxies, obtained a linear $N_\rmn{X}-\rmn{SFR}$ relation.
Finally,  in interpreting the $N_\rmn{X}-\rmn{SFR}$ and $L_\rmn{X}-\rmn{SFR}$ relations one should keep in mind that for the XLF slope of $\approx 1.6$ the number of sources is defined by faint sources, near the detection limit whereas their total luminosity -- by the (sometimes just a few) brightest sources.    

\section{Collective luminosity of HMXBs}
\label{sec:ltot}

The collective X-ray luminosities of X-ray binaries in galaxies from our primary sample were computed as the sum of luminosities of point sources detected above the corresponding sensitivity limit $L_{\rmn{lim}}$. The latter was defined as the luminosity level corresponding to a value of the incompleteness function $K(L)=0.6$, in the regions selected as described in Sect. \ref{sec:spatial}.
As \textit{Chandra} observations of different galaxies achieved different sensitivity, these luminosities need to be transformed to the same threshold, which was chosen at $L_{\rmn{X}} = 10^{36}$ erg/s, i.e. the luminosities presented below are an estimate of the combined luminosity of X-ray binaries brighter that this threshold. The luminosity  of "missing" unresolved sources was computed  by integrating the luminosity function which differential slope was fixed at the average value of $\gamma=1.58$, determined in section \ref{sec:xlf}. The normalization was determined for each galaxy  individually, from the number of sources detected above its sensitivity limit, corrected for incompleteness and subtracting the CXB contribution as described in section \ref{sec:xlf}.  Finally the luminosities were corrected for the contribution of the background AGN to the resolved sources:
\begin{equation}
\label{eq:lx}
L^{\rmn{XRB}}_{0.5-8}=\int_{10^{36}}^{L_{\rmn{lim}}}L\frac{\rmn{d}N}{\rmn{d}L}\rmn{d}L+\sum_{L_{i}\geq L_{\rmn{lim}}} \frac{L_i}{K(L_{i})}-4\pi D^2 F_\rmn{CXB}
\end{equation}
where $F_\rmn{CXB}$ is the predicted total flux of CXB sources brighter than the flux limit corresponding to $L_\rmn{lim}$ computed from the $\log N-\log S$ of \citet{2008MNRAS.388.1205G} and $D$ is the distance to the galaxy.
The CXB contribution did not exceed $\sim 35-40\%$ of the total luminosity of galaxies, in accord with the procedure used to define the analysis regions (Section \ref{sec:prof_cxb}). For the reasons described in Section \ref{sec:lmxb_cxb}, we did not attempt to subtract the contribution of LMXBs. 

\begin{figure*}
\centering
\begin{minipage}{165mm}
\hbox
{
\includegraphics[width=0.5\linewidth]{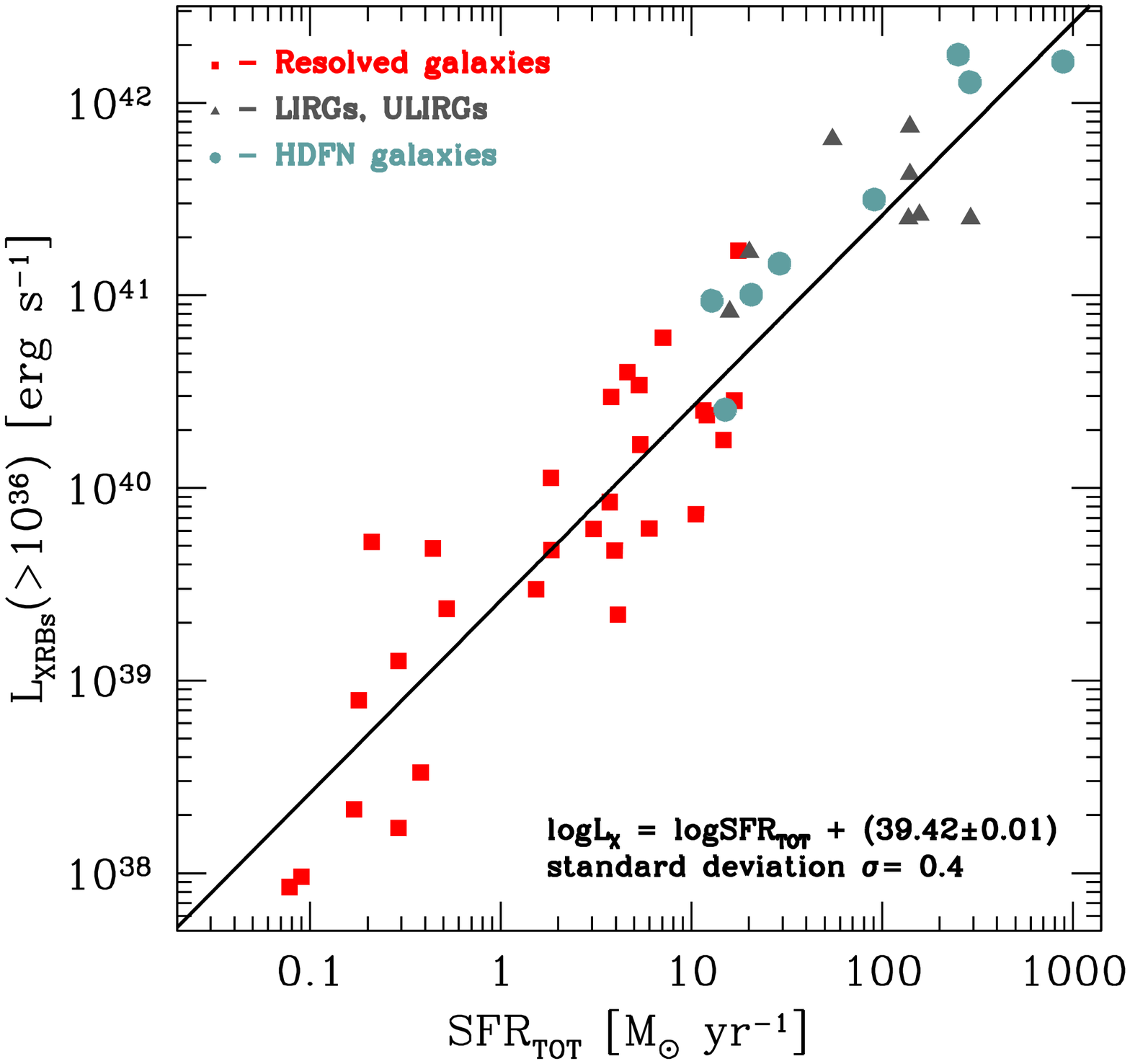}
\includegraphics[width=0.5\linewidth]{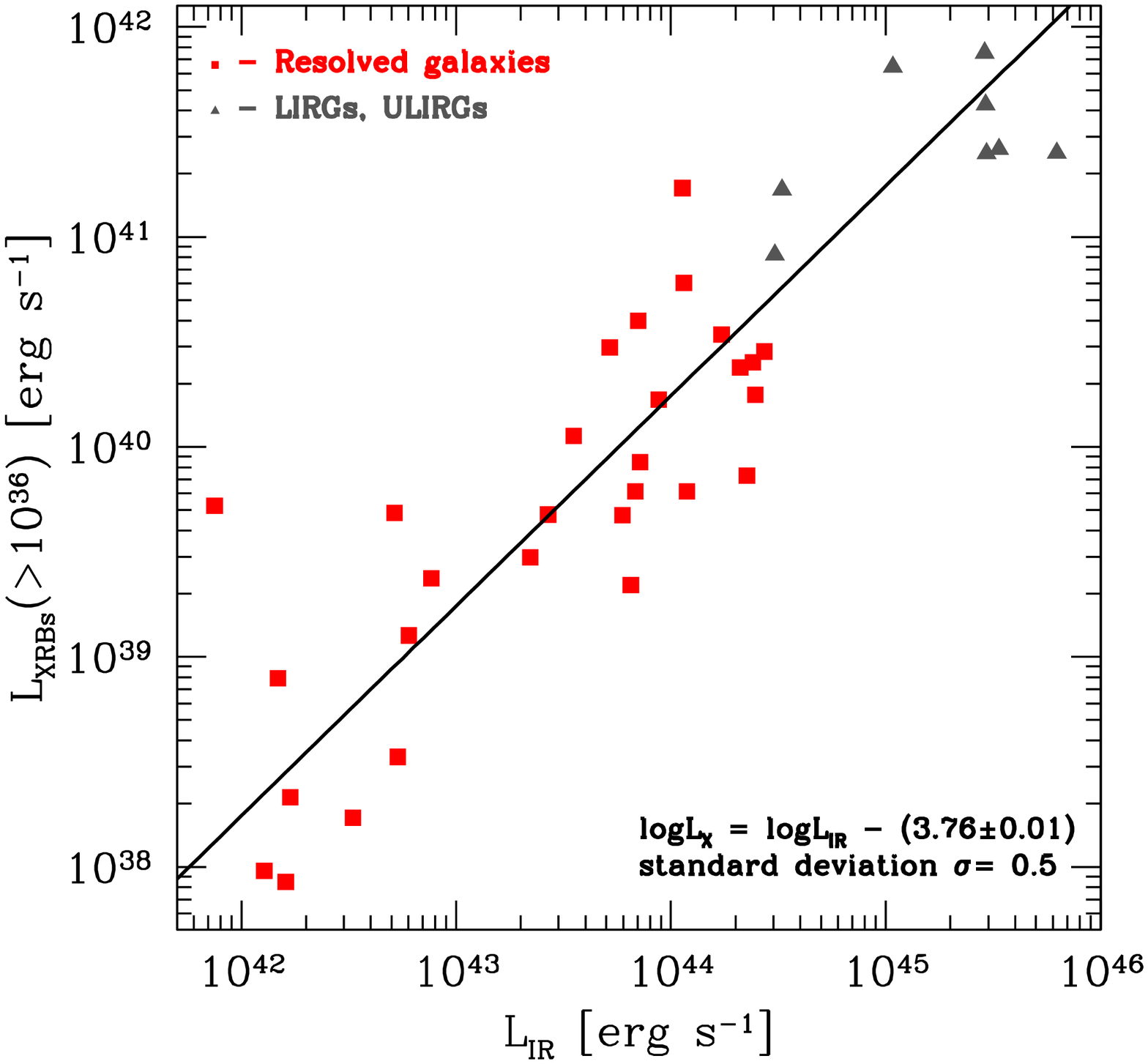}
}
\caption{The $L_{\rmn{X}}-\rmn{SFR}$ ({\em left}) and  $L_{\rmn{X}}-L_{\rmn{IR}}$  ({\em right}) relations. In both panels the squares are galaxies from the resolved sample, the triangles and filled circles are galaxies from the unresolved high-SFR sample. The solid lines show the best-fitting scaling relations, obtained using only resolved galaxies and given by eqs. (\ref{eq:lx_sfr_linear}) and (\ref{eq:lx_lir_linear}) respectively.} 
\label{fig:lx_sfr}
\end{minipage}
\end{figure*}

The obtained $L_{\rmn{XRB}}-\rmn{SFR}$ relation is shown in the left-hand panel of Fig. \ref{fig:lx_sfr}. We also show in this plot the data for unresolved galaxies from the high-SFR sample. The latter, however were not used for the fits because of the contribution of unresolved emission, as discussed in section \ref{sec:remark}. A least-squares fit to the data with the relation $\log L_\rmn{X}=\log K + \beta\log \rmn{SFR}$ yielded a value of slope consistent with  unity: $\beta = 1.01 \pm 0.11$. 
We therefore fixed the slope at the unity and repeated the fit to obtain  the best-fitting linear relation:
\begin{equation}
\label{eq:lx_sfr_linear}
L^{\rmn{XRB}}_{0.5-8\,\rmn{keV}} (\rmn{erg}\,\rmn{s}^{-1}) = 2.61\times 10^{39}\,\rmn{SFR}\,(M_{\odot}\,\rmn{yr}^{-1})
\end{equation}
The dispersion around this relation is $\sigma=0.43$ dex.

To facilitate comparison with previous results, we also studied the $L_\rmn{X}-L_\rmn{IR}$ relation (right-hand panel of Fig. \ref{fig:lx_sfr}). The linear fit of this relation gives the following result:
\begin{equation}
\label{eq:lx_lir_linear}
L^{\rmn{XRB}}_{0.5-8\,\rmn{keV}} (\rmn{erg}\,\rmn{s}^{-1}) = 1.75\times 10^{-4}\,L_{\rmn{IR}}\,(\rmn{erg}\,\rmn{s}^{-1})
\end{equation}
where $L_{\rmn{IR}}$ is the total ($8-1000\,\umu \rmn{m}$) infrared luminosity. The dispersion around the best-fit is $\sigma=0.54$ dex, i.e. notably worse than for the $L_\rmn{X}-\rmn{SFR}$ relation with SFR based on combined UV+IR data. 

\subsection{Unresolved galaxies}

As it is obvious from Fig.\ref{fig:lx_sfr},  unresolved galaxies are distributed along the extension of the $L_{\rmn{X}}-\rmn{SFR}$ relation for resolved galaxies, albeit somewhat higher. Quantitatively, a linear fit to the $L_{\rmn{X}}-\rmn{SFR}$ relation using unresolved galaxies only yielded the scale factor of $\approx 3.7\times 10^{39}$ erg/s/(M$_\odot$/yr), i.e. by a factor of $\sim 1.5$ higher than in eq.(\ref{eq:lx_sfr_linear}). This result is to be expected because the luminosities of unresolved galaxies, by definition, include the unresolved emission component, which is excluded in the case of the galaxies from our main sample. Although the detailed study of unresolved emission is beyond the scope of this paper and will be fully addressed in the Paper II (in preparation), we made a preliminary estimate of its amplitude in several galaxies from the primary sample, uniformly distributed along the SFR axis. We found that unresolved emission contributes $\sim 20-60\%$ to the total X-ray luminosity of these galaxies.  This is consistent with the higher normalization in the scaling relation for unresolved galaxies, obtained above.

\subsection{Comparison with previous studies}

To do the comparison, we converted the results of the earlier studies to the 0.5--8 keV band  assuming an absorbed power law with $\Gamma=2.0$ and $n_{H}=3\times 10^{21}$ cm$^{-2}$.

The first systematic studies of the $L_{\rmn{X}}-\rmn{SFR}$ scaling relations based on the data of modern X-ray satellites were performed by \citet{2002AJ....124.2351B}, \citet{2003A&A...399...39R} and \citet{2003MNRAS.339..793G}. 

In, probably the first, Chandra-based publication on this subject,
\citet{2002AJ....124.2351B} studied the correlation between X-ray and 1.4 GHz
luminosity of faint X-ray sources in the Hubble Deep Filed North. Using their eq.(5)
we find that in the SFR range of interest their $L_\rmn{X}/\rmn{SFR}$ ratio exceeds by a factor
of $\sim 3-4$ the value obtained in the present work. As this is also inconsistent
by about same factor with the results of many other studies, we will not attempt to
investigate the origin of this discrepancy.
We note however, that the scale factor of their 1.4 GHz -based SFR determination is
similar to the one used in this paper, which produced consistent results for the
CDF-N galaxies.

\citet{2003A&A...399...39R} obtained tight correlation between X-ray, radio and FRI luminosities  for a sample of nearby galaxies using X-ray data from ASCA and BeppoSAX satellites. Using their eqs (8) and (12), converting X-ray luminosities from 0.5--10.0 to 0.5--8.0 keV band (a factor of 1.11)  and FIR  ($42.5-122.5\,\umu$m) luminosity to total IR ($8-1000\,\umu$m) band using a conversion factor of 1.7, we obtain the relation $L_{0.5-8\,\rmn{keV}}\approx 2.4\cdot 10^{-4}\times L_\rmn{IR}$. The scale factor in this relation is by $\approx 1.36$ larger than in eq.(\ref{eq:lx_lir_linear}), which is consistent with the fact that \citet{2003A&A...399...39R} analyzed the total luminosities of galaxies, including unresolved emission. We conclude that their $L_{\rmn{X}}-L_{\rmn{IR}}$ relation is consistent with the one derived in this paper. 

To derive $L_{\rmn{X}}-\rmn{SFR}$ relation we use their eqs.(14) and (15) and obtain $L_{0.5-8\,\rmn{keV}}\approx 8.6\cdot 10^{39}\times \rmn{SFR}(>5 M_\odot)$ where $\rmn{SFR}(>5 M_\odot)$ is the formation rate of massive stars. For Salpeter IMF $\rmn{SFR}(0.1-100M_\odot)=5.5\cdot \rmn{SFR}(>5 M_\odot)$, we thus obtain $L_{0.5-8\,\rmn{keV}}\approx 1.6\cdot 10^{39}\times \rmn{SFR}(0.1-100 M_\odot)$. This is close but not identical to the relation derived in the present paper, eq.(\ref{eq:lx_sfr_linear}), especially considering that it applies to the total X-ray luminosity of the galaxy, not only bright X-ray binaries. 
As the underlying scaling relations between X-ray and IR luminosities are fully consistent with our results, the factor of $\sim 2-2.5$ discrepancy in the $L_\rmn{X}-\rmn{SFR}$ relation must be  due the different method of the SFR estimation used in the two studies.

\citet{2003MNRAS.339..793G} used {\it Chandra} observations of nearby (i.e. spatially resolved) star-forming galaxies and  derived a relation between luminosity of compact sources and SFR. Using their eq.(21) and converting the luminosity to 0.5--8 keV band (conversion factor of 1.28) we obtain: 
$L_{0.5-8\,\rmn{keV}}\approx 8.5\cdot 10^{39}\times \rmn{SFR}(M>5M_\odot)$, i.e. identical to that obtained by  \citet{2003A&A...399...39R}. \citet{2005A&A...431..597S} recomputed the scale in this relation with account for the updated calibration of IR and radio SFR indicators \citep{2003ApJ...586..794B}. Using their scale factor we derive $L_{0.5-8\,\rmn{keV}}\approx 2.8 \cdot 10^{39}\times \rmn{SFR}(0.1-100M_\odot)$, in good agreement with eq.(\ref{eq:lx_sfr_linear}).

As a historical note we mention that the SFR values determined from the IR-based calibration  used by \citet{2003A&A...399...39R} and \citet{2003MNRAS.339..793G} were incorrectly attributed by the authors to the formation rate of massive stars. This was a result of some confusion that existed in the SFR determination-related literature at the time of those publications and, in particular, by inconsistent definitions of radio- and IR-based SFR proxies.  This error  could potentially result in a factor of $\sim 5.5$ overestimate of the scale in the $L_\rmn{X}-\rmn{SFR}$ relation. However, it was compensated by the usage of FIR instead of total IR luminosity in  the \citet{1998ARA&A..36..189K} relation and  the scale factors somewhat different from the presently accepted values. As a result, the final $L_\rmn{X}-\rmn{SFR}$ relations derived by these authors are in a reasonable agreement with the scaling relations obtained in this paper.

Among more recent publications on this subject, we compare with the results of \citet{2007A&A...463..481P}, \citet{2008ApJ...682.1020K} and \citet{2010ApJ...724..559L}.

\citet{2007A&A...463..481P} investigated the relation between the collective hard X-ray luminosity of young point sources and the SFR derived from infrared luminosity. Using their eq.(9) and converting the 2--10 keV luminosity to the 0.5--8 keV band (a factor of 1.28) we obtain: $L_{0.5-8\,\rmn{keV}}\approx 0.96\cdot 10^{39}\times \rmn{SFR}(0.1-100M_\odot)$. This scale factor is significantly, by a factor of $\sim 2-2.5$ lower than derived in all other studies, including this one. The reason for this discrepancy is, in our opinion, in an overestimated contribution of the old population to the X-ray emission of star-forming galaxies. 

\citet{2008ApJ...682.1020K} examined the X-ray point-source population of two galaxies with high SFRs (NGC~4194 and NGC~7541) for which they found  $L_{\rmn{X}}/L_{\rmn{IR}}$ ratio consistent with our average relation, eq. (\ref{eq:lx_lir_linear}). However, their $L_\rmn{X}/\rmn{SFR}$ ratio is by a factor of $\sim 2$ larger than ours, due to a different SFR calibration used. 

Based on {\it Chandra} observations of nearby luminous infrared galaxies (LIRGs) \citet{2010ApJ...724..559L} studied the relation between 2--10 keV luminosity and SFR. To compare results, we convert their $2-8$ keV luminosity to the $0.5-8$ keV band and obtain the scale factor for the linear $L_\rmn{X}-\rmn{SFR}$ relation (first line in their Table 4) of 
$L_\rmn{X}/\rmn{SFR} \approx 2.2\cdot 10^{39} \rmn{erg}\,\rmn{s}^{-1} (M_{\odot}\,\rmn{yr}^{-1})^{-1}$, that is in a good agreement with our result.

\subsection{Effects of statistics of small numbers in the $L_\rmn{X}-\rmn{SFR}$ relation} 
\label{sec:stat}

For the power law luminosity function with a slope of 1.6,  the  $L_\rmn{X}-\rmn{SFR}$ relation  is expected to be non-linear in its low-SFR part, due to the effects of statistics of small numbers (Gilfanov, Grimm, \& Sunyaev 2004a, 2004b).
These effects are important when considering the collective luminosity of a population of discrete sources and are a consequence of an asymmetric probability distribution for the sum of luminosities of compact sources in the limit of their small number. In the case of HMXBs in star-forming galaxies, they make the $L_\rmn{X}-\rmn{SFR}$ relation to appear non-linear in the low SFR limit. This effect is purely statistical in its nature, in the sense that it affects the most likely value of the luminosity of a randomly chosen galaxy, whereas the relation between average luminosity of many galaxies and SFR is linear in the entire range of star-formation rates. This is achieved due to extended tail of the probability distribution $p(L_{tot})$ towards large $L_{tot}$.  In the sample of galaxies from Gilfanov, Grimm, \& Sunyaev (2004a, 2004b)
such a non-linear dependence was clearly observed at $\rmn{SFR}\la 5~M_\odot$/yr.

Despite nearly identical XLF of compact sources obtained in this paper, the $L_\rmn{X}-\rmn{SFR}$ relation appears to be linear in the entire SFR regime. In an attempt to find the reason of such behavior, we investigated the distribution of the luminosity of the brightest source in a galaxy and found that for a fraction of low-SFR galaxies it is higher than predicted based on the average XLF (the predicted value computed according to formulae from \citet{2004MNRAS.347L..57G}). As XLF slopes of low-SFR galaxies are consistent with the rest of the sample (Fig.\ref{fig:slopes}) and the contribution of background AGN is negligible in the bright end of the XLF, we conclude that this may be a consequence of the observer bias, introduced by selection of galaxies -- targets for {\it Chandra} observations (see Sect. \ref{sec:obs_bias}). For example, the proposers  may have predominantly selected galaxies which  were known  to have enhanced level of X-ray emission, e.g. based on the ROSAT all-sky survey.  This issue will be investigated in more detail in a follow-up publication.

\begin{figure*}
\begin{center}
\hbox
{
\includegraphics[width=58mm]{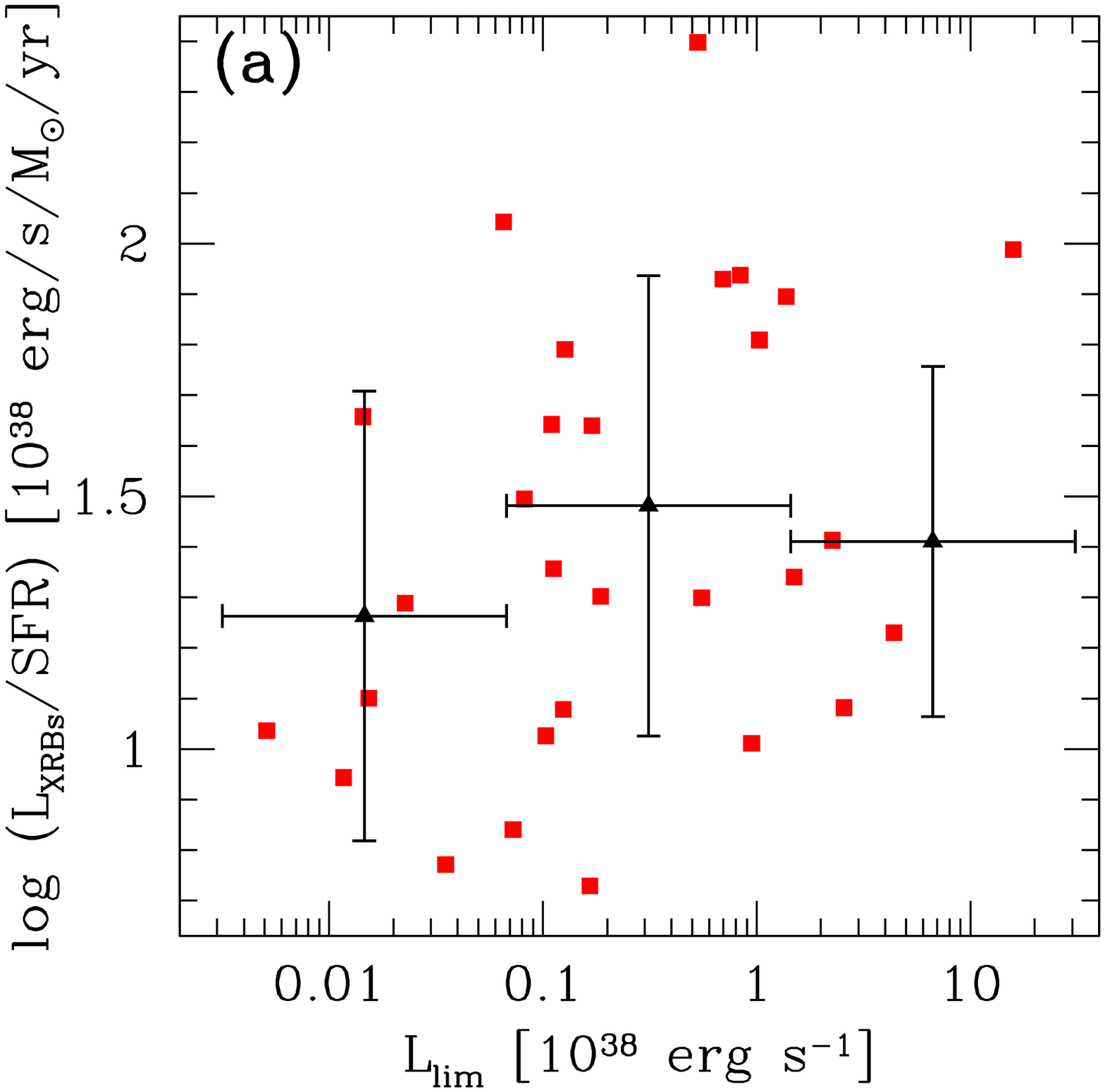}
\includegraphics[width=58mm]{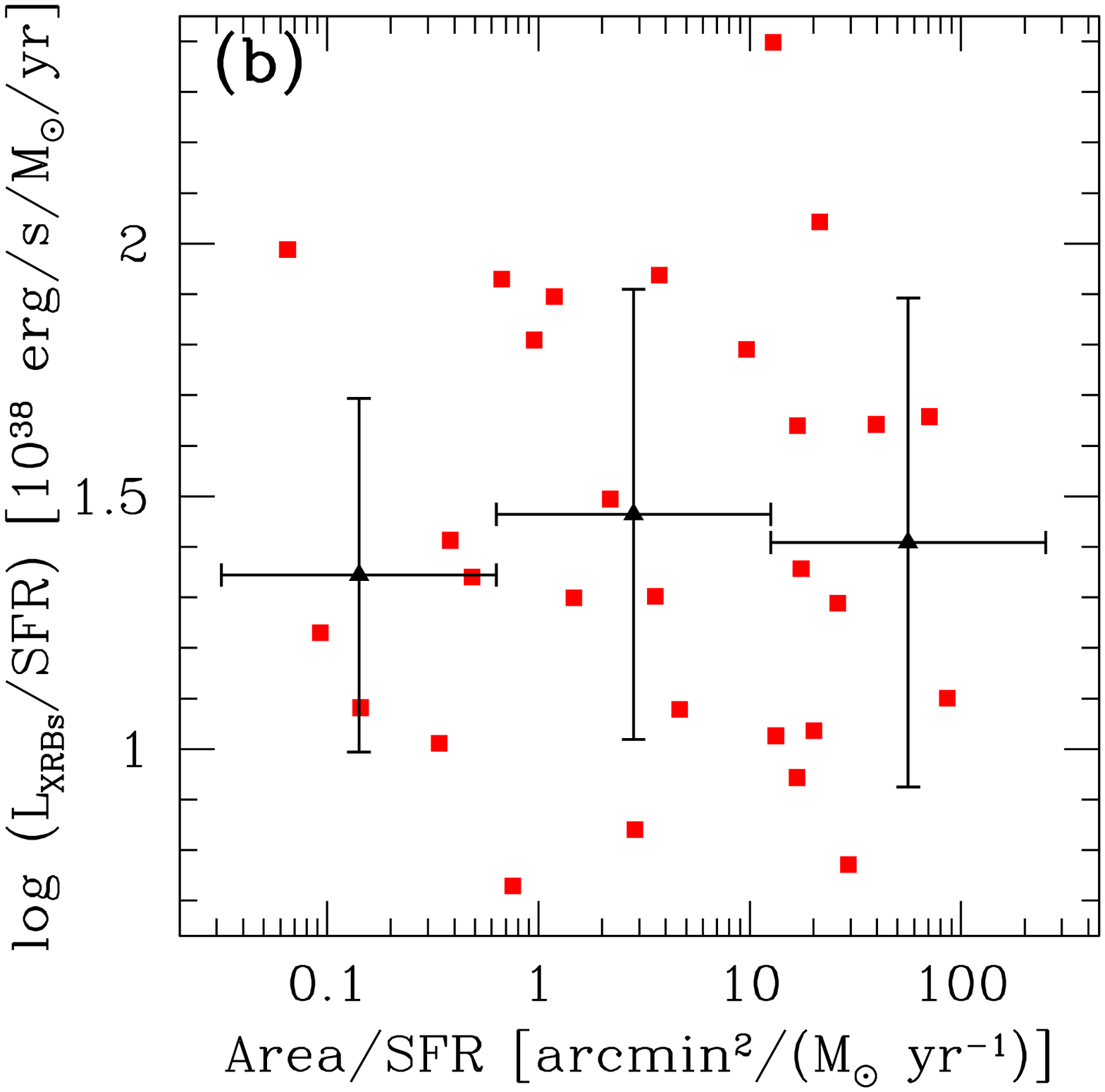}
\includegraphics[width=58mm]{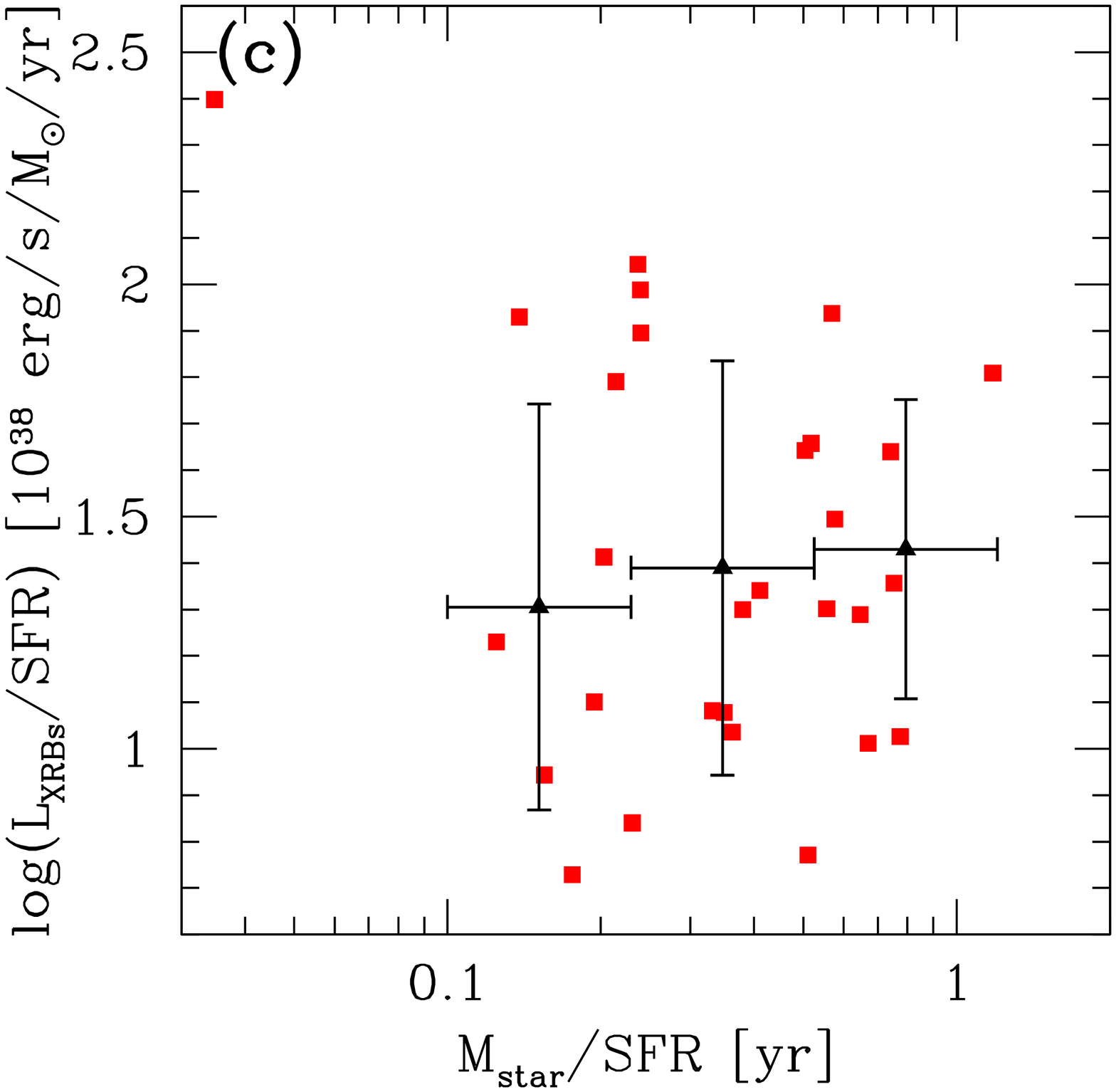}
}
{
\includegraphics[width=58mm]{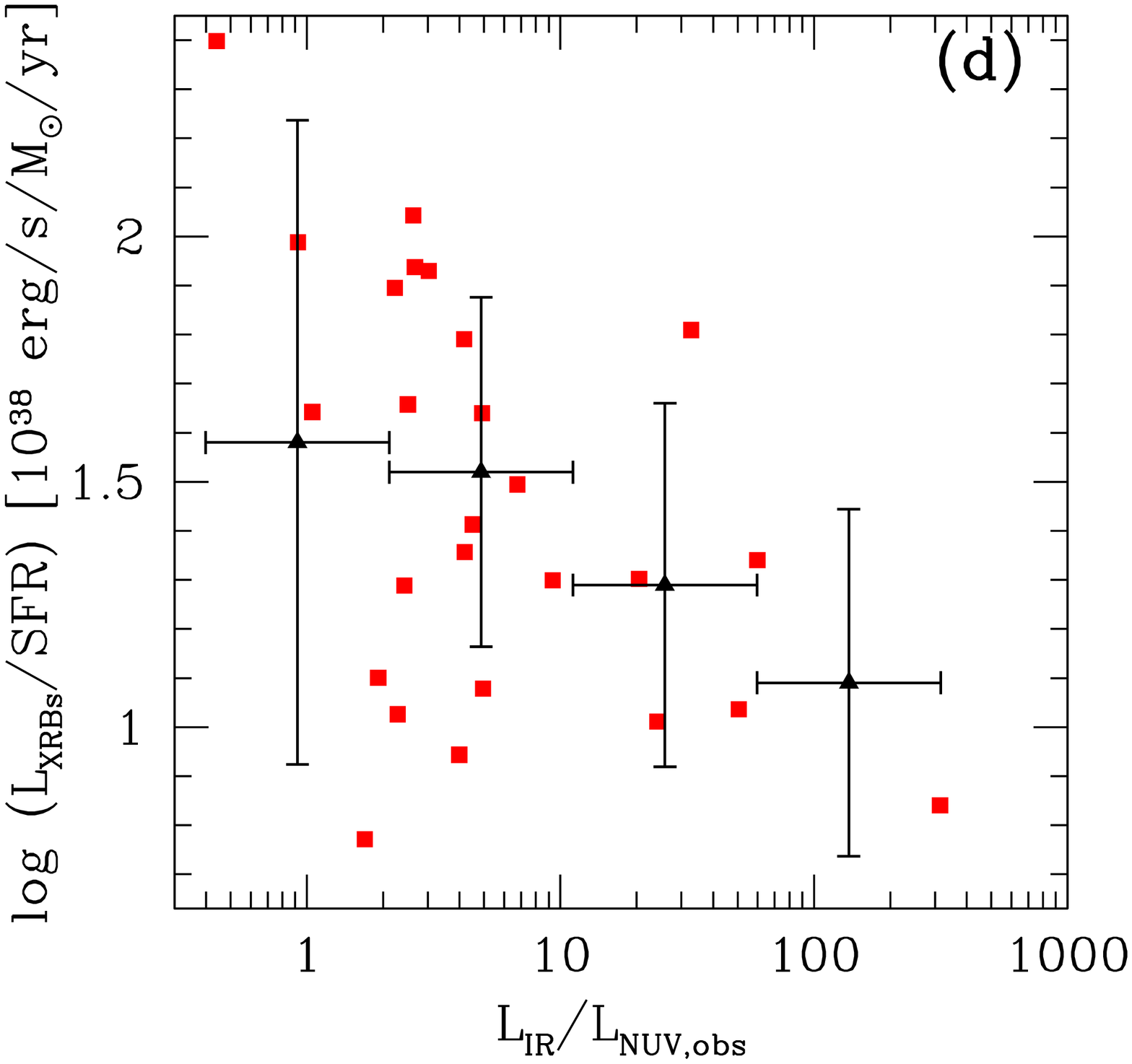}
\includegraphics[width=58mm]{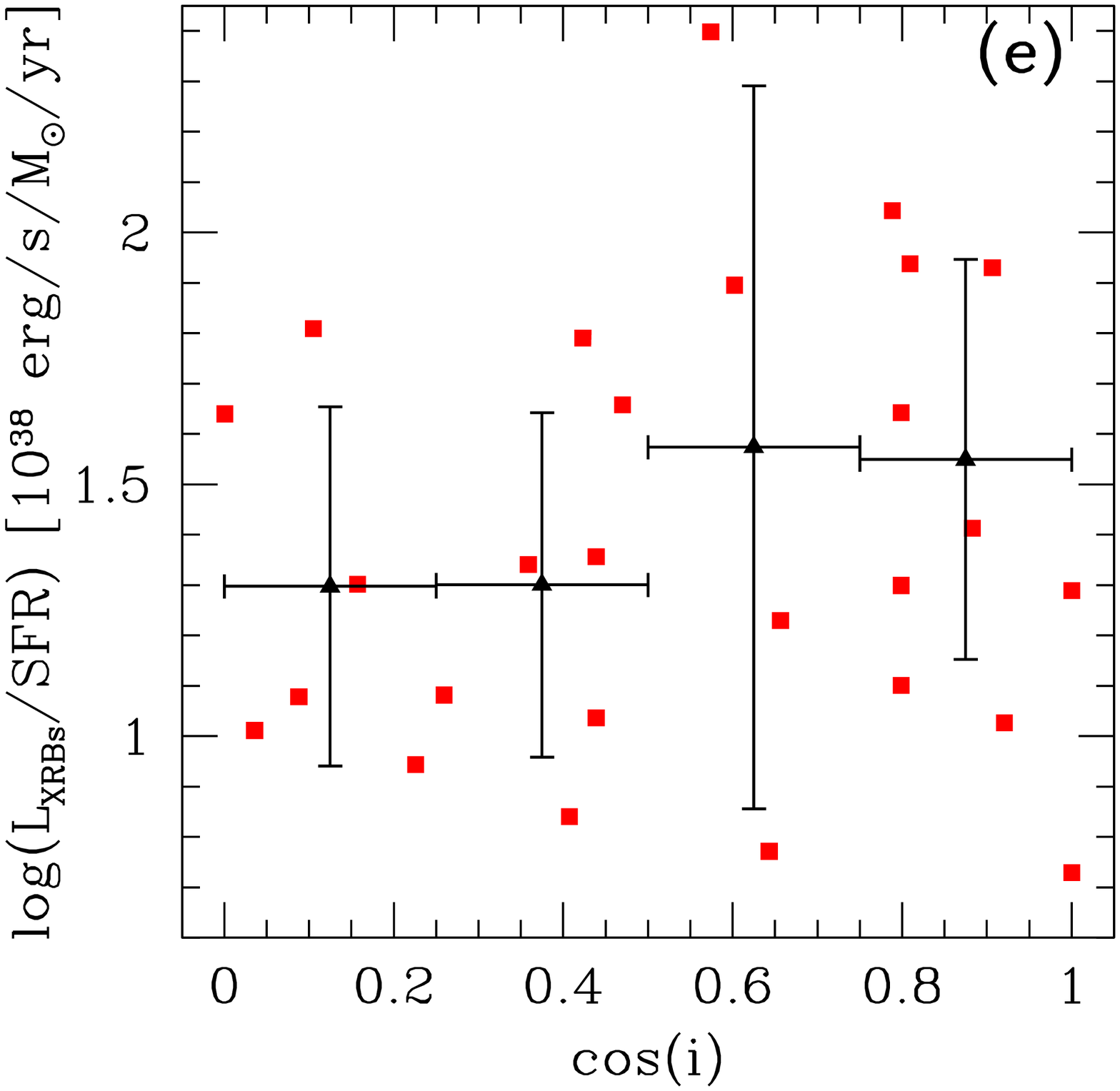}
\includegraphics[width=58mm]{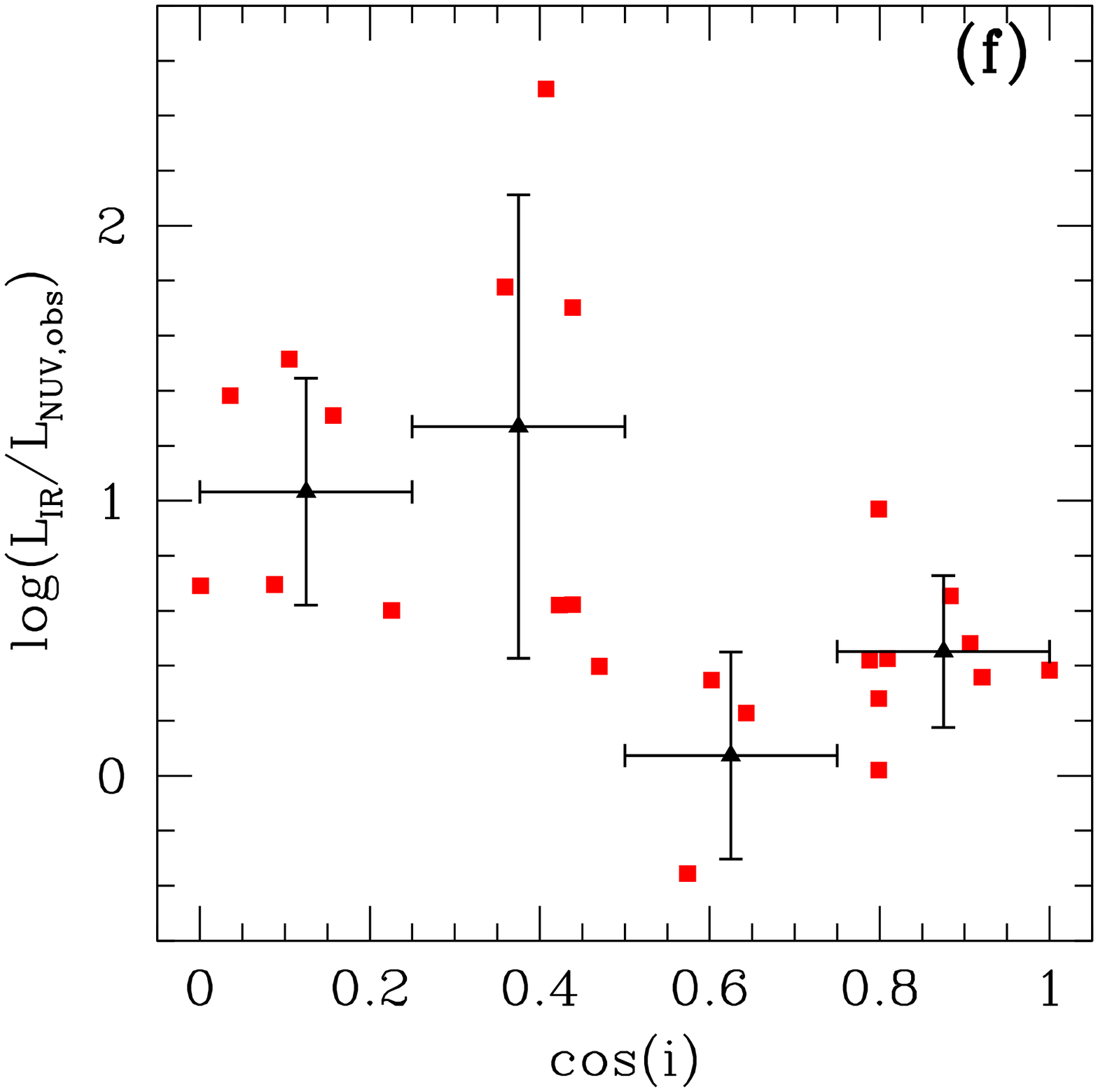}
}
\caption{Dependence of the $\log(L_\rmn{X}/\rmn{SFR})$ on quantities characterizing different potentially contaminating factors: the point source detection sensitivity of Chandra observations (a), the fractional contribution of  CXB sources (b) and LMXBs (c), the  dust attenuation characterized by $L_{\rmn{IR}}/L_{\rmn{NUV}}$ ratio (d) and the inclination of the galaxy. The last panel (e) shows the relation between the dust attenuation and inclination. Each point represents a galaxy from the primary sample. The crosses show average values in bins; the bin extent is indicated by the horizontal error bar, the vertical error bar shows the rms of points with respect to the average value. The contribution of CXB sources is subtracted. See Sect \ref{sec:dispersion} for details.}
\label{fig:dispersion_study}
\end{center}
\end{figure*}

\subsection{Dispersion in the $L_{\rmn{XRB}}-\rmn{SFR}$ relation}
\label{sec:dispersion}

Despite of the special effort to  produce a sample with a homogeneous set of multi-wavelength measurements and to minimize contamination by CXB sources and LMXBs, the resulting $L_{\rmn{XRB}}-\rmn{SFR}$ relation bears a rather large dispersion of $\sim 0.4$ dex rms.
In order to investigate the origin of this dispersion we searched for correlation of the $L_{\rmn{XRB}}/\rmn{SFR}$ ratio with different quantities characterizing importance of various potential contaminating factors. These included the sensitivity limit of \textit{Chandra} data $L_{\rmn{lim}}$, contribution of CXB and LMXB sources characterized by the ratio of the solid angle of the studied region and enclosed stellar mass to the SFR, dust attenuation characterized by the $L_{\rmn{IR}}/L_{\rmn{UV}}$ ratio and inclination of the galaxy $i$. The results are shown in Fig. \ref{fig:dispersion_study}. 

We analyzed these data using the Spearman's rank correlation test and  found no statistically significant correlations. The results of these tests are presented below with $r_{S}$ being the correlation coefficient and $P$ being the probability under the null hypothesis that the variables are unrelated. We found no correlation with the point source detection sensitivity  ($r_{S} = 0.33$, $P = 8\%$). Therefore we can exclude that the extrapolation of X-ray luminosity above $10^{36}$ erg/s is the main reason for the dispersion around the  $L_{\rmn{XRB}}-\rmn{SFR}$ relation. 
There is no evidence for strong contamination by CXB sources ($r_{S} = -0.03, P = 86\%$) or LMXBs ($r_{S} = -0.03, P = 88\%$).  This proofs that the spatial analysis, performed in order to minimize the contribution of CXB and LMXB sources to the compact source population, yielded sensible results. It is further confirmed by repeating the previous test on the data without CXB subtraction: $r_{S} = -0.01, P = 96\%$. 
We used the ratio of IR to NUV luminosity $L_{\rmn{IR}}/L_{\rmn{NUV}}$ to characterize the dust attenuation (higher the ratio, higher the absorption) and found virtually no correlation ($r_{S} = -0.35, P = 7\%$). Neither $L_\rmn{X}/\rmn{SFR}$ correlates with the inclinations of galaxies ($r_{S} = 0.09, P = 65\%$). To verify this approach, we analyzed the dependence between $L_{\rmn{IR}}/L_{\rmn{NUV}}$ ratio and inclination of the galaxy and found a strong anti-correlation ($r_{S} = -0.59, P = 0.2\%$), as expected. 
We note that in order to suppress the $0.5-8$ keV luminosity by a factor of 10 a large column density of $\sim 5\times 10^{23}$ cm$^{-2}$ is needed.

\begin{figure}
\begin{center}
\includegraphics[width=1.0\linewidth]{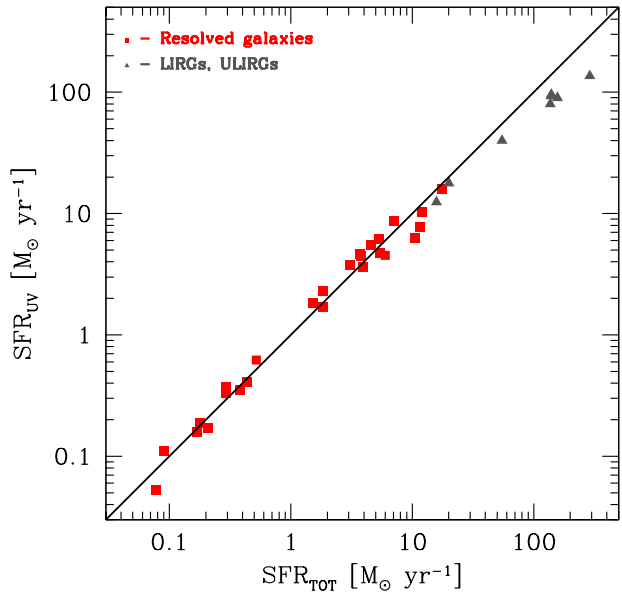}
\caption{Comparison of the SFR estimations obtained by different methods. The x-axis is the SFR used in the paper, computed as described in Section \ref{sec:sfr}. The y-axis values are computed from the UV luminosity corrected for the dust attenuation using the method of \citet{2005ApJ...619L..51B}. The straight line is $y=x$ relations. The dispersion of points around this relation is $rms=0.12$.} 
\label{fig:sfrcheck}
\end{center}
\end{figure}

\begin{figure}
\begin{center}
\includegraphics[width=1.0\linewidth]{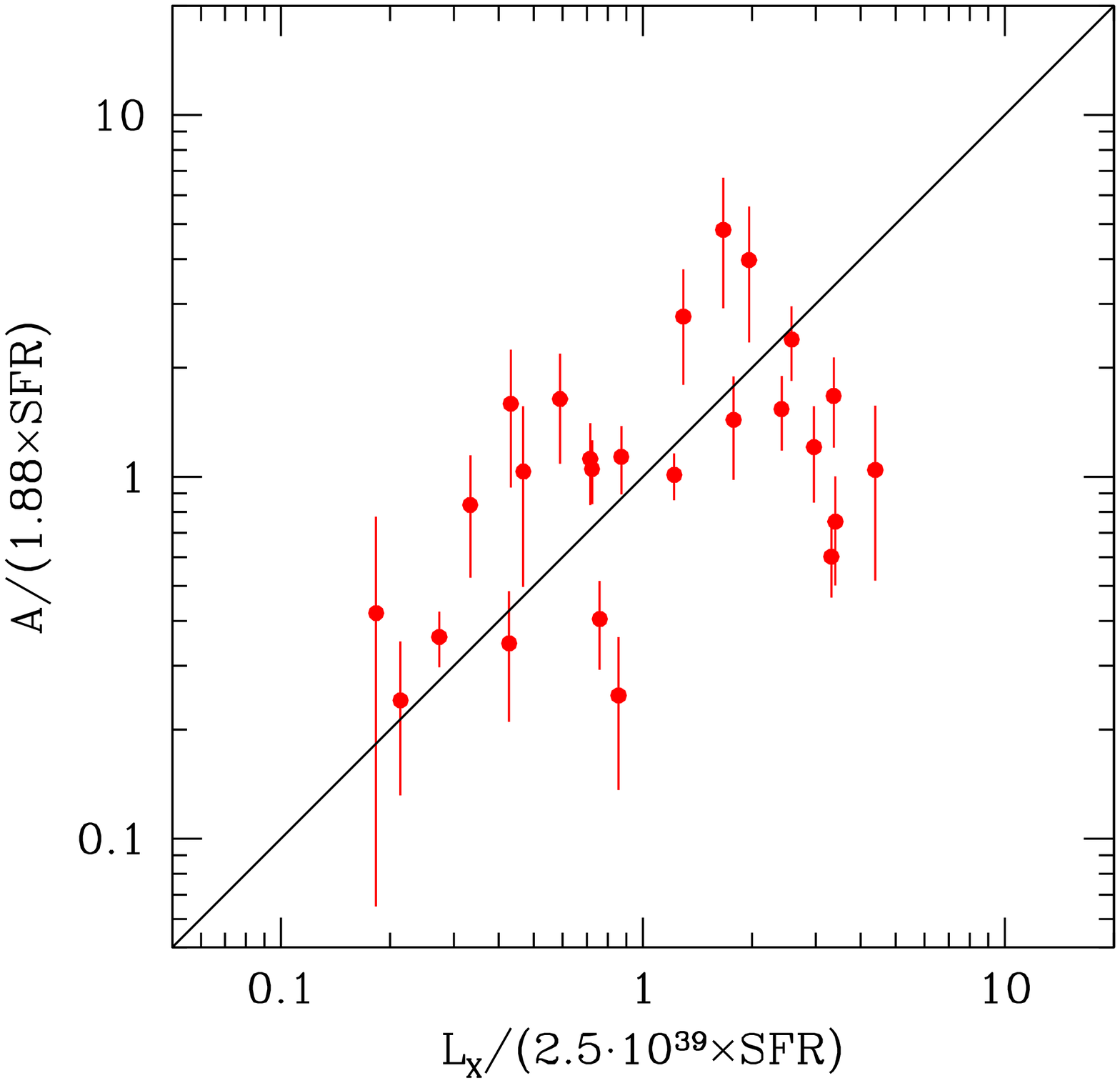}
\caption{The XLF normalization vs. total luminosity of X-ray binaries. Both quantities are normalized to the SFR with the normalization coefficients equal to the best fit values in the corresponding scaling relations. Only galaxies having $\ge 5$ sources above their threshold luminosities are shown. This plot illustrates the lack of one-to-one correspondence between the deviations from the average dependences  in the number of sources and in the  total luminosity.} 
\label{fig:lx_nx}
\end{center}
\end{figure}

We also tried to probe the amplitude of intrinsic uncertainties of the SFR determination. To this end we computed the UV based SFR estimates corrected for the attenuation effects following the method of \citet{2005ApJ...619L..51B}, and compared with the SFR values used in this paper (Fig.\ref{fig:sfrcheck}). We found a good consistency between the SFR estimations obtained by the two methods with $rms=0.12$ dex, i.e. three times smaller than the rms of the $L_X-SFR$ relation. This suggests that uncertainties of the SFR determination are not the main source of the scatter in the $L_\rmn{X}-\rmn{SFR}$ relation.

\section{Discussion}

We have found rather large scatter around the average $L_\rmn{X}-\rmn{SFR}$ relation $rms\approx 0.4$ dex (Fig.\ref{fig:lx_sfr}). 
As discussed in the Section \ref{sec:dispersion} the scatter is unlikely to be caused by any of the obvious contaminating factors, such as difference in the sensitivity limits of observations of different galaxies, pollution by CXB sources or LMXBs. This proofs that the criteria used for selecting the sample and the method for choosing the regions for the analysis yielded the desired result. On the other hand it suggests that the observed scatter has physical origin. 

\begin{figure}
\begin{center}
\includegraphics[width=1.0\linewidth]{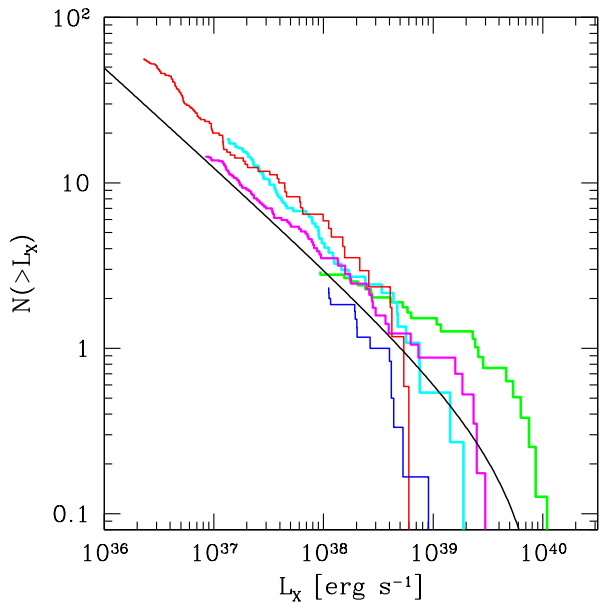}
\caption{Several representative XLFs, normalized to the SFR. In the order of increasing luminosity of the brightest sources:  M101, NGC~3079, M51, Antennae, NGC~3310. The solid line shows a power law with average normalization of $A=1.88$,  slope $\gamma=1.6$  and a cut-off at $L_\rmn{cut}=10^{40}$ erg/s. The XLFs of NGC~3079 and NGC~3310 are, correspondingly, the steepest and the flattest in the sample.} 
\label{fig:xlfs}
\end{center}
\end{figure}

Although similar scatter is found in the SFR-dependence of the XLF normalization, $rms\approx 0.34$ dex (Fig.\ref{fig:xlfnorm}), there is no precise one-to-one correspondence between the deviations of these two quantities from their best fit relations, as illustrated by Fig.\ref{fig:lx_nx}. At the same time the XLF slopes appear to be consistent, within statistical errors, with all having the same (or close) value  (Fig.\ref{fig:slopes}, \ref{fig:xlfs}). These two facts do not contradict to each other. Indeed, for a power law with $\gamma=1.6$ the XLF normalization and, largely, its slope are determined by the more numerous faint sources, whereas the total luminosity is determined by the (often few) brightest sources in the galaxy.  Thus, we conclude that the scatter in the $L_\rmn{X}-\rmn{SFR}$ relation is caused by both variations in the XLF normalization and fluctuations in the number and luminosity of brightest sources in the galaxy (cf. Fig.\ref{fig:xlfs}), the variations of the XLF slopes playing less significant role. 

Metallicity is one of the primary quantities that may affect the population of HMXBs, as it exerts an equally great effect on HMXB creation and emission \citep[e.g.][]{1978ApJ...221L..37C}. It has a rather substantial impact on the strength of stellar winds, thus on the luminosity of individual HMXBs, making it larger for larger amount of metals present in the companion star. Star formation at lower metallicity is expected to produce more massive stars as a result of the low cooling rate. This effect would result in a higher number of compact objects in binary systems. This simple theoretical argument is supported by the recently found evidence that the number of ultra-luminous X-ray sources  is inversely proportional to the metallicity of their host galaxy \citep{2009MNRAS.400..677Z, 2010MNRAS.408..234M}. However, this is in contrasts with population synthesis calculations of  \citet{2010ApJ...725.1984L}, who found that the existence of massive single star BHs at low metallicities does not directly correlate to the existence of more massive compact black holes in binary systems. 
Despite this, they find the same inverse relationship between the number of ultra-luminous X-Ray
sources and metallicity, described in \citet{2010MNRAS.408..234M}. 
In an attempt to advance this issue observationally, we collected from the literature a rather heterogeneous set of the metallicity measurements for a relatively small subset of galaxies, but could not find any meaningful trends in the data.  This, however, should not be considered as an ultimate result, more systematic studies are obviously needed in order to make a reliable conclusion regarding the role of metallicity (work in progress).

The synthetic XLF produced by \citet{2010ApJ...725.1984L} has a high luminosity turnover at approximately $10^{40}$ erg/s, close to the observed value, though the slope of the latter XLF is somewhat harder ($1.16$) than observed. The possible reason for the latter may be that \citet{2010ApJ...725.1984L}  do not consider variable HMXBs such as Be-HMXBs which provide much of the lower-luminosity population (T. Linden, private communication).

Another important factor is the recent star-formation history of the host galaxy. Indeed,  as more massive stars have shorter life times, and, on the other hand, have stronger winds one may expect that the very luminous HMXBs are short-lived. This effect may result in the dependence of the X-ray luminosity of the galaxy on the recent star-formation activity, which is not characterized precisely the SFR value determined from the IR and/or UV data, thus introducing scatter in the $L_\rmn{X}-\rmn{SFR}$ relation.

In addition to the more fundamental reasons discussed above, variability of  the brightest X-ray sources may also contribute to the scatter in the $L_\rmn{X}-\rmn{SFR}$ relation. The X-ray sources with $\log L_\rmn{X} \ga 39$ (usually classified as ultra-luminous X-ray sources) likely harbor a black hole and may be subject to the disk instability resulting in transient events characterized by luminosity variations of several orders of magnitude \citep{1996ARA&A..34..607T}. In addition, spectral state changes may lead to the luminosity changes from the factor of a few upto $\sim$order of magnitude. Their effect on the total luminosity will be especially important in galaxies with $\rmn{SFR}\la 10$ M$_\odot$/yr, containing only a few such bright sources \citep{2004MNRAS.351.1365G}.

\subsection{XLF of high-mass X-ray binaries and ULXs.}

Based on a sample of $\sim 700$ compact sources, we have produced the average luminosity distribution of HMXBs in galaxies  (Fig.\ref{fig:dxlf}). This is a significant improvement in the statistical accuracy as compared with the previous result of \citet{2003MNRAS.339..793G}. We have also improved in terms of the errors introduced by the distance uncertainties, thanks to the systematic compilation of distances for galaxies which became available in the NED. Thirdly, we made a special effort to minimize and accurately subtract the contribution of background AGN. Although we could not in the same way remove the contribution of LMXBs,  because of the strong age dependence of the LMXB--stellar mass  scaling relations, the method of construction makes us confident that our sample is relatively free from the LMXB contamination. This is definitely true in the high luminosity end, $\log L_\rmn{X}\ga 38.5-39$. 

Despite all these improvements our average HMXB XLF is entirely consistent with the one obtained by \citet{2003MNRAS.339..793G}. Indeed, the values of slope agree to the accuracy of $\sim 0.03$ and the values of the high luminosity break are consistent within (admittedly rather large) statistical uncertainties. 

Similar to \citet{2003MNRAS.339..793G}, we did not find any statistically significant features in the XLF near the critical Eddington luminosity of a neutron star or a stellar mass black hole. The XLF follows a single slope power in the entire luminosity range, upto  $\log L_\rmn{X}\sim 40$, where it breaks. Taken at the face value, the break corresponds to the Eddington luminosity of a $\sim 50-100$ M$_\odot$ black hole. 
We note that the statistical  accuracy of the XLF, grouped into broad, $\Delta \log L\sim 0.5$, luminosity bins, is $\la 13\%$ and $\la 22\%$ in the $\log L_\rmn{X}\sim 38-39$ and   $\log L_\rmn{X}\sim 39-40$ luminosity ranges (Fig.\ref{fig:dxlf2}). This further strengthens the proposition of \citet{2003MNRAS.339..793G} that the bulk of the population of what is called the ultra-luminous X-ray sources (conventionally defined as sources with $\log L_\rmn{X}\ga 39$) are a high luminosity end of the population of the ordinary HMXBs and most likely harbor stellar mass black holes rather than exotic intermediate mass objects. The very high luminosity end of the distribution, $\log L_\rmn{X}\ga 40$, may, however, be associated with the intermediate mass black holes.  

The absence of features in the luminosity function of HMXBs is still puzzling.  The inaccuracies in  distances to the galaxies in our sample may still contribute to some smearing of the break at the NS Eddington limit, but their role must be significantly smaller than in the XLF of \citet{2003MNRAS.339..793G}. 
Indeed, the average distance uncertainty for the galaxies from our sample is $\sim 15\%$ (Table \ref{table:sample1}) which translates to luminosity  uncertainty of $\Delta \log(L_X)\approx  0.12$. This corresponds to the half of the bin width for the XLF shown in Fig.\ref{fig:dxlf} (and $\la 1/4$ of that for the one shown in Fig.\ref{fig:dxlf}) and can not have  any major effect on the XLF shape.
Taken at the face value, this result may suggest that there is no sharp limit on the luminosity of an accreting compact object at the Eddington value, as discussed, for example, by \citet{2002A&A...391..923G, 2003MNRAS.339..793G}. This does not explain, however, why the nearly an order of magnitude difference in the  specific  frequencies of neutron stars and black holes does not reveal itself in the XLF. Indeed, in the entire luminosity range the XLF slope is consistent with the slope of the distribution of the mass transfer rates in HMXBs, predicted from the standard Salpeter IMF for massive stars \citep{2003AstL...29..372P}.

Answers to these questions could be obtained if one could  disentangle the contributions of NS and BH binaries and separate their XLFs. Direct measurements of the mass of the compact object in a large number of binaries in external galaxies is not feasible. Therefore indirect methods will have to be utilized, for example based on the X-ray spectral properties of the binaries. For example, one of such methods, based on the X-ray colors, is being developed by Farrell et al (in preparation).

\section{Implications for the theory of binary evolution} 

\subsection{Specific frequency of X-ray bright compact objects in HMXBs}
\label{sec:fx_hmxb}

With the knowledge of the relation between the number of HMXBs and SFR, we can estimate the fraction of compact objects that went through  an X-ray active phase  powered by accretion of matter from a massive donor star in a binary system.

According to eq. (\ref{eq:dlf}), the number of HMXBs with luminosity higher than $10^{35}$ erg/s is:
\begin{equation}
N_{\rmn{HMXB}}(>10^{35} \rmn{erg}\,\rmn{s}^{-1}) \approx 135 \times \rmn{SFR} 
\label{eq:Nhmxb_lowL}
\end{equation}
On the other hand, the number of HMXBs at any given time is:
\begin{equation}
N_{\rmn{HMXB}}\sim \dot{N}_{co}\, \sum_k f_{\rmn{X},k}\, \tau_{\rmn{X},k} 
\sim \dot{N}_{co} \, f_{\rmn{X}} \,  \bar{\tau}_{\rmn{X}}
\label{eq:Nhmxb_anytime}
\end{equation}
where $\dot{N}_{co}\approx \dot{N_{\star}}(M>8\,M_{\odot})$ is the birth rate of compact objects (both NS and BH), approximately equal to the birth rate of massive stars $\dot{N}_{\star}(M>8\,M_{\odot})$. For a Salpeter IMF, we obtain:
\begin{equation}
\dot{N}_{*}(M>8\,M_{\odot}) \approx 7.4 \cdot 10^{-3} \times \rmn{SFR} 
\label{eq:birth_rate_ns_bh}
\end{equation}

The summation in eq.(\ref{eq:Nhmxb_anytime}) is done over different types of HMXBs (wind-fed and Roche lobe overflow systems with a supergiant companion, Be/X binaries etc), $f_{\rmn{X},k}$ is the fraction of compact objects (computed with respect to their total number in the galaxy) which become X-ray sources due to accretion in HMXBs of the $k-$type, $\tau_{\rmn{X},k}$ is the duration of their X-ray active phase. 
The $f_{\rmn{X}}=\sum_k f_{\rmn{X},k}$  is the total fraction of X-ray bright compact objects  and  $\bar{\tau}_{\rmn{X}}$ is the average life time of X-ray sources:
\begin{equation}
\bar{\tau}_{\rmn{X}}=\frac{\sum_k f_{\rmn{X},k} \tau_{X,k}} {\sum_k f_{\rmn{X},k}}
\label{eq:tau}
\end{equation} 
From binary evolution calculations,  $\tau_{\rmn{X}} \sim 10^{4}\,\rmn{yr}$ for supergiant systems and  $\tau_{\rmn{X}} \sim 10^{5}\,\rmn{yr}$ for Be/X binaries. Majority of known high-mass X-ray binaries are Be/X systems, especially in Magellanic Clouds, where only a few supergiant systems are known \citep{2006A&A...455.1165L}. The fraction of the latter is larger in the Milky Way, where about 29 out of 114 HMXBs (are suspected to) have supergiant donors.  If these proportions are representative for star-forming  galaxies in general, the summation in eq.(\ref{eq:Nhmxb_anytime}) and (\ref{eq:tau}) is dominated by Be/X systems and appropriate scaling for $\bar{\tau}_{\rmn{X}}$ should be $\sim 10^5$ yrs. 

Combining equations \ref{eq:Nhmxb_lowL}, \ref{eq:Nhmxb_anytime} and \ref{eq:birth_rate_ns_bh}, we obtain:
\begin{equation}
f_{\rmn{X}} \sim 0.18 \times \bigg(\frac{\bar{\tau}_X}{0.1\,\rmn{Myr}}\bigg)^{-1}
\label{eq:fraction_ns_bh_hmxb}
\end{equation}
This quantity represents the fraction of compact objects that on the time scale of $\sim 10^2$ Myrs after their formation \citep[e.g.][]{2007AstL...33..437S} become X-ray sources with $L_\rmn{X}> 10^{35}$ erg/s, powered by accretion from a massive donor star in a high-mass X-ray binary. 

It is interesting to note, that if the HMXB XLF continues towards lower luminosities with $\sim$ same or similar  slope, it may be possible to constrain $\bar{\tau}_{\rmn{X}}$ based on the specific frequency of X-ray sources in star-forming galaxies. Indeed,  if XLF from eq.(\ref{eq:dlf}) can be extrapolated  down to $L_\rmn{X}\sim 10^{34}$ erg/s, as suggested by \citet{2005MNRAS.362..879S}, the coefficient in eq.(\ref{eq:fraction_ns_bh_hmxb}) becomes 0.72. As $f_{\rmn{X}}$ can not exceed the binary fraction $f_{bin}\sim 0.5$, the X-ray life-time of Be/X systems is $\bar{\tau}_{\rmn{X}}\la 0.07\, (f_{bin}/0.5)$ Myrs.

\subsection{Bright sources -- Black hole systems}
\label{sec:fx_bh}

The derivation of eq.(\ref{eq:fraction_ns_bh_hmxb}) assumed that  $f_{\rmn{X}}$ and $\bar{\tau}_{\rmn{X}}$ do not depend on the mass of the compact object which assumption is likely to be incorrect.  However, as we used a rather low luminosity limit of $10^{35}$ erg/s in eq.(\ref{eq:Nhmxb_lowL}), the $N_\rmn{HMXB}$  is dominated by faint Be/X binaries, which harbor neutron stars. Therefore the $f_{\rmn{X}}$ derived above characterizes neutron stars and is likely to be insensitive to the contribution of black holes.
As contributions of BH and NS systems to the $N_\rmn{HMXB}$ in eq.(\ref{eq:Nhmxb_lowL}) can not be separated, we can not estimate $f_{\rmn{X}}$ for BH binaries of all luminosities separately. However, this can be done for bright sources, with luminosity significantly exceeding the Eddington limit for the neutron star. Using $10^{39}$ erg/s as the luminosity limit, we obtain:
\begin{equation}
N_{\rmn{HMXB}}(>10^{39} \rmn{erg}\,\rmn{s}^{-1}) \approx 0.49 \times \rmn{SFR} 
\label{eq:Nhmxb_highL}
\end{equation}
As we consider luminous sources, the $\dot{N}_{co}$ in eq.(\ref{eq:Nhmxb_anytime}) is now the birth rate of black holes, which for Salpeter IMF is:
\begin{equation}
\dot{N_{*}}(M>25\,M_{\odot}) \approx 1.4 \cdot 10^{-3} \times \rmn{SFR}  
\end{equation}
The resulting fraction of stellar-mass BH that became  luminous X-ray sources in HMXB is:
\begin{equation}
f^{bright}_{\rmn{X}} \sim 3.5\cdot 10^{-2} \times \bigg(\frac{\tau_{\rmn{HMXB}}}{10^{4} \,\rmn{yr}}\bigg)^{-1} 
\label{eq:fraction_bh}
\end{equation}
In the latter equation we took into account that  bright HMXBs shining near the Eddington limit typically contain supergiant donors, therefore their characteristic life times are shorter than for Be/X systems and are $\sim 10^4$ yrs.

Finally we note that in the estimates in sections \ref{sec:fx_hmxb} and \ref{sec:fx_bh} we intentionally used Salpeter mass function rather than a more realistic IMF.   Indeed, the UV and IR SFR proxies used in this paper rely on the emission from massive, $M\ga 5M_\odot$, stars. The star-formation rates derived by these proxies are then converted to the full $0.1-100 M\odot$ range assuming Salpeter IMF  \citep[e.g.][]{1998ARA&A..36..189K}. Therefore, for the sake of consistency, same IMF should be used in the estimates of the numbers of compact objects in eq.(\ref{eq:birth_rate_ns_bh}) and (\ref{eq:fraction_bh}).    In the remaining part of this section a more correct approach is to use a realistic IMF, correctly predicting the frequencies of low mass stars, e.g. the one suggested by \citep{2002Sci...295...82K}.

\subsection{Specific frequency of X-ray bright compact objects in LMXBs}
\label{sec:fx_lmxb}

For comparison, we obtain a similar estimate for the fraction of compact objects that become X-ray sources powered by accretion from a low-mass donor star in an  LMXB. According to \citet{2004MNRAS.349..146G}, the number of LMXBs with luminosity higher than $10^{35}$ erg/s is:
\begin{equation}
N_{\rmn{LMXB}}(>10^{35} \rmn{erg}\,\rmn{s}^{-1}) \approx 50 \times M_*/(10^{10} M_\odot)
\label{eq:Nlmxb_lowL}
\end{equation}
Assuming stationarity:
\begin{equation}
N_{\rmn{LMXB}}\sim N_{co} \, f_{\rmn{X}}
\label{eq:nlmxb}
\end{equation}
Strictly speaking, stationarity assumption does not apply to bright LMXBs. Indeed, the life time of a source with  $L_{\rmn{X}}=10^{38}$ erg/s ($\dot{M}\sim 10^{-8}$ M$_\odot$/yr) powered by accretion from a $\sim 1$ M$_\odot$ star can not exceed $\sim 10^8$ yrs. However, as in the case of HMXBs, the $N_\rmn{LMXB}$ in eq.(\ref{eq:Nlmxb_lowL}) is dominated by faint sources, for which the stationarity assumption may apply.
Assuming a Kroupa IMF (slope of $\gamma=1.3$ for $M<0.5 M_\odot$,  $\gamma=2.3$ for $0.5<M<1 M_\odot$ and $\gamma=2.7$  for $M>1 M_\odot$  \citep{2002Sci...295...82K}), the number of neutron stars and black holes in  such a galaxy would be:
\begin{equation}
N_{co}\approx N_{*}(M>8\,M_{\odot}) \approx 5\cdot 10^{-3} \times M_*
\label{eq:nr_ns_bh}
\end{equation}
Combining eqs. (\ref{eq:Nlmxb_lowL})-(\ref{eq:nr_ns_bh}) we obtain:
\begin{equation}
f_{\rmn{X}} \sim 10^{-6} 
\label{eq:fraction_ns_bh_lmxb}
\end{equation}
According to the derivation, this number takes into account only LMXBs in bright state and ignores compact objects in transient systems which are in the quiescent state.

It is interesting to note that the $f_{\rmn{X}}$ for LMXBs  is by $\sim 5$ orders of magnitude smaller than for HMXBs. This is another manifestation of the fact that LMXBs are extremely rare objects and may be explained by the high probability of disruption of the binary system with a low mass companion in the course of the supernova explosion (e.g. Verbunt \& van den Heuvel 1995).

\subsection{Constrains on the mass-ratio distribution in binaries}

As not every compact object is in a binary and not every compact object in a binary has a suitable combination of binary parameters to become an HMXB, $f_{\rmn{X}}$ derived in section \ref{sec:fx_hmxb} obeys the condition: 
\begin{equation}
f_{\rmn{X}}\le f_{bin}(m_2\ga 5 M_\odot) 
\label{eq:fx_fm}
\end{equation}
where $f_{bin}(m_2\ga 5 M_\odot)$ is the fraction of binaries with compact object, having a massive companion. This inequality can be used obtain interesting constrains on the mass distribution of the secondary in massive binaries.

Firstly, we can exclude the  possibility that the mass of the secondary in a binary system obeys a Kroupa mass function. Indeed, for Kroupa MF, the fraction of stars more massive than $5M_\odot$ is $f(m> 5 M_\odot) \approx 5.5\cdot 10^{-3}$ by number, i.e.  is by a factor of $\sim 30$ to $\sim 100$ smaller than $f_{\rmn{X}}$, in a strong contradiction to eq.(\ref{eq:fx_fm}). Considering a more general power law mass distribution $\psi(m)\propto m^{-\gamma}$, and assuming $f_{\rmn{X}}=0.2$ we obtain from eq.(\ref{eq:fx_fm}) a constrain $\gamma<0.3$, i.e. much flatter than the high mass end of Kroupa or Salpeter mass function.

If the mass transfer in the course of the binary evolution does not significantly change the mass of the secondary, the above result strongly excludes the possibility that the masses of stars in a massive binary are drawn independently  from the IMF. 
Assuming  further that number of binaries obeys the distribution in the form 
\begin{equation}
dN\propto \psi(m_1) dm_1\, f(q) dq
\end{equation}
where  $m_1$ is the mass of the primary, with the mass distribution $\psi(m_1)$ and $f(q)$ is the distribution of the mass ratio $q=m_2/m_1$, we write for the fraction of system with the secondary mass $m_2>5M_\odot$ among all binaries with the primary mass $m_1>8M_\odot$ (i.e. those producing a compact object):
\begin{equation}
f_{bin}(m_2\ga 5 M_\odot) =\frac{\int_8^{100} dm_1 \psi(m_1) \int_{5/m_1}^1 f(q) dq}{\int_{0.1}^{100} \psi(m) dm}
\end{equation}
For the Kroupa IMF and $f(q)=1$ the above equation gives  $f_{bin}(m_2\ga 5 M_\odot) \approx 0.6$, i.e. consistent with $f_{\rmn{X}}\sim 0.2$.

\section{Conclusions}
Based on a homogeneous  set of X-ray, infrared and ultraviolet observations from \textit{Chandra}, Spitzer and GALEX archives, we studied the properties of populations of high-mass X-ray binaries in a sample of 29 nearby star-forming galaxies and their relation with the SFR. We took a special care to minimize the contamination by LMXBs, background AGN and to control the incompleteness of the Chandra source lists.

Our main findings  can be summarized as follows:

\begin{inparaenum}[i\upshape)]
\item We have detected $1057$ compact X-ray sources, of which $\sim 300$ are expected to be background AGN.  The majority of remaining $\sim 700$ sources are  young systems associated with star-formation in the host galaxy. Based on their high X-ray luminosities and analogy with the X-ray populations in the Milky Way and few other very nearby galaxies, we conclude that they are high-mass X-ray binaries, powered by accretion of matter from a massive donor star in a binary system.
 
\item The shape of their luminosity distribution is similar in different galaxies with the power low indexes having $rms=0.25$ with respect to the average value of $\approx 1.6$.  The XLF normalization, on the contrary,  show significantly larger dispersion with the  $rms=0.34$ dex around the $A\propto \rmn{SFR}$ law. 

\item Combining the data of all galaxies we produce the average XLF. Despite its by far better statistical accuracy the obtained XLF is entirely consistent with the one produced by \citet{2003MNRAS.339..793G} based on a much smaller sample of galaxies and much fewer sources. The HMXB XLF has a slope of $\gamma\approx 1.6$ in the broad luminosity range $\log L_\rmn{X}\sim 35-40$ and shows a moderately significant  evidence for the high luminosity break or cut-off at $\log L_\rmn{X}\approx 40$. Similar to Grimm et al, we did not find any statistically significant features at the Eddington luminosity limits of neutron stars or stellar mass black holes. 

\item HMXBs are a good tracer of the recent star formation activity in the host galaxy. Their collective luminosity  is proportional to the SFR:
\begin{equation}
L^{\rmn{XRB}}_{0.5-8\,\rmn{keV}} (\rmn{erg}\,\rmn{s}^{-1}) = 2.61\times 10^{39}\,\rmn{SFR}\,(M_{\odot}\,\rmn{yr}^{-1})
\end{equation}
The $rms$ of points around this relation is 0.4 dex.  The observed dispersion is unlikely to be caused by any of the obvious contaminating factors such as CXB or LMXB sources and is likely to have a physical origin.

\item The collective luminosity of X-ray binaries is well correlated with the infrared luminosity of the host galaxy:
\begin{equation}
L^{\rmn{XRB}}_{0.5-8\,\rmn{keV}} (\rmn{erg}\,\rmn{s}^{-1}) = 1.75\times 10^{-4}\,L_{\rmn{IR}}\,(\rmn{erg}\,\rmn{s}^{-1})
\end{equation}
where $L_{\rmn{IR}}$ is the total ($8-1000\,\umu \rmn{m}$) infrared luminosity. The dispersion around the best-fitting relation is $\sigma=0.5$ dex. 

\item  The number of bright  HMXBs scales with the SFR of the host galaxy:
\begin{equation}
N_{\rmn{XRB}}(>10^{38} \rmn{erg}\,\rmn{s}^{-1}) = 3.22  \times \rmn{SFR}\,(M_{\odot}\, \rmn{yr}^{-1})
\end{equation}
with the dispersion of $rms=0.34$ dex.

\item The fraction of compact objects which were X-ray bright at least once in their lifetime, powered by accretion from a massive star in a binary is $f_{\rmn{X}}\sim 0.2$. This constrains the mass distribution of the secondary in massive binaries. For an independent mass distribution of the secondary, the power law index must be flatter than $0.3$.  In particular, an independent  mass distribution of a Kroupa or Salpeter type is strongly excluded. Assuming that the masses of components in a binary are not independent, our results are consistent with the flat mass ratio distribution. 

The fraction of compact objects, X-ray active in LMXBs, on the contrary, is small, $f_{\rmn{X}}\sim 10^{-6}$, demonstrating that LMXBs are extremely rare objects. This result is in line with the conclusions of the binary evolution theory (e.g. Verbunt \& van den Heuvel 1995).

\end{inparaenum}

\section*{Acknowledgments}

SM is grateful to Rasmus Voss, for support and tools for X-ray photometry and XLF incompleteness correction, and to Pepi Fabbiano, Sean Farrell, Peter Jonker, Ginevra Trinchieri, Anna Wolter, Luca Zampieri for discussions.
This research made use of \textit{Chandra} archival data and software provided by the \textit{Chandra} X-ray Center (CXC) in the application package CIAO. This research has made use of SAOImage DS9, developed by Smithsonian Astrophysical Observatory.
The \textit{Spitzer Space Telescope} is operated by the Jet Propulsion Laboratory, California Institute of Technology, under contract with the NASA. \textit{GALEX} is a NASA Small Explorer, launched in 2003 April.
This publication makes use of data products from Two Micron All Sky Survey, which is a joint project of the University of Massachusetts and the Infrared Processing and Analysis Center/California Institute of Technology, funded by the NASA and the National Science Foundation.
This research has made use of the NASA/IPAC Extragalactic Database (NED) which is operated by the Jet Propulsion Laboratory, California Institute of Technology, under contract with the National Aeronautics and Space Administration.

\appendix

\section[]{Catalogue of HMXB in nearby star-forming galaxies}

\begin{table*}
\begin{minipage}{180mm}
\caption{List of compact sources detected within the $D25$ ellipse in all galaxies of the resolved sample.}
\label{table:catalogue}
\begin{tabular}{@{}l c c c c c c c c c c @{}}
\hline
Galaxy & ID & CXO name & $\alpha_{\rmn{J2000}}$ & $\delta_{\rmn{J2000}}$ & d & $S$ & $\sigma_S$ & $\log F_{\rmn{X}}$ & $\log L_{\rmn{X}}$ & location \\
  & &  & (deg) & (deg) & (arcsec) & (cts) & (cts) & ($\rmn{erg}\,\rmn{cm}^{-2}\,\rmn{s}^{-1}$) & ($\rmn{erg}\,\rmn{s}^{-1}$)  & flag\\
  (1) & (2) & (3) & (4) & (5) & (6) & (7) & (8) & (9) & (10)  & (11)\\
\hline
\hline
NGC~0278 &1  & CXOU J005204.3+473304 & 13.01831 & 47.55115 &   2.5 &  41.7 &   9.6 & -14.42 & 37.80 &     3\\
&2  & CXOU J005204.3+473258 & 13.01823 & 47.54965 &   3.1 &  28.7 &   8.4 & -14.58 & 37.64 &     3\\
&3  & CXOU J005204.6+473305 & 13.01957 & 47.55146 &   5.2 &  50.1 &  10.0 & -14.34 & 37.88 &     3\\
&4  & CXOU J005203.8+473309 & 13.01596 & 47.55265 &   9.1 &  62.5 &  10.3 & -14.20 & 38.02 &     1\\
&5  & CXOU J005203.4+473254 & 13.01429 & 47.54855 &  11.4 &  74.9 &  11.0 & -14.15 & 38.07 &     1\\
&6  & CXOU J005204.4+473314 & 13.01835 & 47.55404 &  12.8 &  30.5 &   8.0 & -14.54 & 37.68 &     1\\
&7  & CXOU J005203.3+473312 & 13.01410 & 47.55333 &  13.9 &  25.1 &   7.4 & -14.62 & 37.60 &     1\\
&8  & CXOU J005205.8+473302 & 13.02457 & 47.55056 &  16.1 &  31.5 &   7.8 & -14.54 & 37.68 &     1\\
&9  & CXOU J005202.7+473307 & 13.01137 & 47.55217 &  17.1 & 1035.4 &  39.1 & -13.00 & 39.22 &     1\\
&10 & CXOU J005205.5+473249 & 13.02299 & 47.54716 &  17.2 &  11.5 &   6.7 & -14.98 & 37.25 &     1\\
&11 & CXOU J005202.1+473316 & 13.00896 & 47.55446 &  26.1 &  16.3 &   6.3 & -14.79 & 37.43 &     1\\
&12 & CXOU J005205.9+473321 & 13.02485 & 47.55610 &  26.2 &  17.9 &   6.5 & -14.78 & 37.44 &     1\\
&13 & CXOU J005206.0+473347 & 13.02540 & 47.56320 &  49.2 & 114.7 &  13.5 & -13.98 & 38.24 &     1\\
\hline
NGC~0520 &1 & CXO J012435.1+034731 & 21.14660 & 3.79199 &   2.3 & 1100.1 &  37.2 & -12.79 & 40.17 &     1 \\
&2 & CXO J012434.8+034729 & 21.14552 & 3.79162 &   3.6 &  50.6 &  11.6 & -14.13 & 38.84 &     1 \\
&3 & CXO J012435.6+034729 & 21.14865 & 3.79150 &   9.7 &  58.1 &  10.0 & -14.07 & 38.90 &     1\\
&4 & CXO J012434.1+034740 & 21.14237 & 3.79447 &  15.4 &  28.0 &   7.3 & -14.34 & 38.62 &     1\\ 
&5 & CXO J012433.5+034733 & 21.13973 & 3.79259 &  23.0 &  15.0 &   5.7 & -14.63 & 38.34 &     1\\ 
&6 & CXO J012433.5+034748 & 21.13958 & 3.79681 &  28.3 & 465.8 &  24.5 & -13.14 & 39.82 &     1\\ 
&7 & CXOU J012436.5+034659 & 21.15243 & 3.78320 &  40.2 &   9.1 &   4.7 & -14.88 & 38.09 &     1\\ 
&8 & CXO J012437.6+034715 & 21.15702 & 3.78772 &  42.6 &  58.8 &   9.8 & -14.07 & 38.90 &     1\\
&9 & CXO J012432.7+034806 & 21.13660 & 3.80198 &  48.5 &  24.1 &   6.8 & -14.45 & 38.52 &     1\\ 
&10 &CXO J012431.9+034703 & 21.13308 & 3.78437 &  55.1 &  13.5 &   5.4 & -14.70 & 38.26 &     1\\ 
&11 &CXO J012438.7+034742 & 21.16147 & 3.79512 &  56.0 &  16.4 &   5.8 & -14.62 & 38.35 &     1\\ 
&12 &CXO J012438.4+034812 & 21.16015 & 3.80340 &  64.0 &  13.0 &   5.5 & -14.70 & 38.27 &     2\\ 
&13 &CXO J012437.8+034625 & 21.15756 & 3.77382 &  78.6 &  45.9 &   9.1 & -14.18 & 38.79 &     1\\ 
&14 &CXO J012435.5+034604 & 21.14791 & 3.76794 &  88.3 &  20.4 &   6.5 & -14.53 & 38.44 &     2\\ 
&15 &CXOU J012441.5+034607 & 21.17299 & 3.76872 & 128.8 &  12.9 &   5.7 & -14.70 & 38.27 &     2\\ 
\hline
NGC~1313 &1 & CXO J031818.2-663004 & 49.57604 & -66.50116 &  16.8 & 385.1 &  22.5 & -12.86 & 38.45 &     1\\ 
&2 & CXO J031818.8-663001 & 49.57878 & -66.50042 &  18.8 & 304.0 &  20.8 & -12.96 & 38.35 &     1 \\
&3 & CXO J031820.0-662910 & 49.58346 & -66.48642 &  48.8 & 5137.4 &  79.6 & -11.68 & 39.62 &     1 \\
&4 & CXOU J031821.2-662858 & 49.58867 & -66.48286 &  63.6 &  15.0 &   5.7 & -14.26 & 37.05 &     1 \\
&5 & CXO J031805.4-663014 & 49.52287 & -66.50414 &  66.6 & 161.2 &  14.9 & -13.22 & 38.08 &     1 \\
&6 & CXO J031806.3-663038 & 49.52662 & -66.51073 &  73.2 &  64.2 &  10.4 & -13.63 & 37.68 &     1 \\ 
&7 & CXO J031823.7-662834 & 49.59902 & -66.47625 &  91.7 &  19.6 &   6.4 & -14.14 & 37.17 &     1 \\  
&8 & CXOU J031829.5-662841 & 49.62308 & -66.47809 & 108.6 &  13.0 &   5.2 & -14.32 & 36.99 &     1 \\ 
&9 & CXO J031830.5-663135 & 49.62741 & -66.52659 & 134.0 &  25.6 &   6.6 & -14.04 & 37.27 &     2 \\  
&10& CXO J031839.5-662920 & 49.66462 & -66.48898 & 144.3 &  14.2 &   5.6 & -14.29 & 37.02 &     2 \\  
&11& CXO J031818.8-663230 & 49.57863 & -66.54172 & 157.4 &  85.7 &  11.1 & -13.51 & 37.80 &     2 \\  
&12& CXO J031748.7-663043 & 49.45354 & -66.51204 & 170.1 &  32.1 &   7.6 & -13.91 & 37.39 &     2 \\  
&13& CXO J031757.7-663224 & 49.49069 & -66.54032 & 186.8 &  33.1 &   7.6 & -13.91 & 37.40 &     2 \\  
&14& CXO J031747.3-663121 & 49.44733 & -66.52265 & 192.7 & 320.1 &  21.3 & -12.92 & 38.39 &     2 \\  
&15& CXO J031742.5-663123 & 49.42745 & -66.52330 & 219.4 & 132.2 &  14.0 & -13.28 & 38.03 &     2 \\  
&16& CXOU J031735.6-663059 & 49.39851 & -66.51654 & 250.4 &  13.1 &   5.4 & -14.28 & 37.03 &     2 \\ 
\hline
\end{tabular}
Note. This table is presented in its entirety in the electronic version (http://www.mpa-garching.mpg.de/$\sim$mineo). An abbreviated version of the table is shown here for guidance as to its form and content. Full catalogue data includes 11 columns of information for 1057 sources. Meanings and units for all columns have been summarized in detail in Appendix A.
\medskip
\end{minipage}
\end{table*}

We present a catalogue of compact X-ray sources detected within the $D25$ ellipse for galaxies from our primary sample. The full catalogue data includes 1057 sources and is currently available at the link shown in the note to the Table \ref{table:catalogue}. The latter includes only a small fraction of the catalogue. Table \ref{table:catalogue} lists the following information for each compact sources: sequence number (col. 2), \textit{Chandra} X-ray Observatory (CXO) source name (JHHMMSS.s-DDMMSS) (col. 3), right ascension and declination (J2000.0) in degrees (cols. 4 and 5), distance from the galaxy center in arcsec (col. 6), number of source counts after background subtraction and its statistical error computed as described in Sec. 3.2 (cols. 7 and 8), decimal logarithm of X-ray flux and luminosity, measured in the $0.5-8\,\rmn{keV}$ band as described in Sec. 3.2 (cols. 9 and 10), location of the source within the $D25$ (col. 11). The latter indicates whether the point source has been detected in the inner galactic region dominated by HMXB population and selected as described in Sect. \ref{sec:spatial} (flag 1), in the outer region dominated by CXB sources (flag 2) or in the bulge region dominated by LMXB population (flag 3).

\bsp
\label{lastpage}
\end{document}